\documentstyle[epsfig,prb,aps]{revtex}

\begin{document}

\title{\bf Andreev states, supercurrents and interface effects in clean 
SN multilayers}
\author{C. Ciuhu, A. Lodder\footnote{Corresponding author, 
e-mail: {\it alod@nat.vu.nl}}, R.E.S Rotadoy, R.T.W. Koperdraad} 
\address{Faculty of Sciences / Natuurkunde en Sterrenkunde, 
Vrije Universiteit, De Boelelaan 1081,\\ 1081 HV Amsterdam, The Netherlands}
\date{\today}
\maketitle

\begin{abstract}
{
We present results for the local density of states in the 
S and N layers of a SN multilayer, and the supercurrent, based on a 
Green's function formalism, as an extension of previous calculations 
on NS, SNS and SNSNS systems. 
The gap function is determined selfconsistently. 
Our systems are chosen to have a finite transverse width. 
We focus on phenomena which occur at so-called critical transverse 
widths, at which a new transverse mode is starting to contribute. 
It appears, that for an arbitrary width the Andreev approximation (AA), 
which takes into account only Andreev reflection at the SN interfaces, 
works well. 
We show that at a critical width the AA breaks down. 
An exact treatment is required, which considers also ordinary reflections. 
In addition, we study the influence of an interface barrier on the 
coupling between the S-layers. 
}
\end{abstract}

\section{Introduction}
\label{intro}

Starting about two decades ago, the interest of developing devices at a 
very small scale gave rise to a new branch in physics, the mesoscopic 
physics. Both theoretically and experimentally, many interesting 
phenomena were discovered which occur at this scale which lies 
essentially in submicron ranges. 

Many samples are built up out of superconducting (S) and normal metallic 
(N) components in which necessarily SN interfaces and possibly 
point contacts occur. 
It is why a lot of theoretical work was devoted to studying 
different SN configurations. 

The first experimental investigations, using tunneling spectroscopy 
measurements, revealed the fact that the density of states in a normal 
metal connected to a superconductor is modified \cite{mcmillan}. 
McMillan provided a simple tunneling model for the proximity 
effect at SN interfaces, which allows for a solution of the Gor'kov 
equations. Ishii \cite{ishii} and Furusaki \cite{furusaki1,furusaki} 
extendend his work to include Andreev reflections. 

Recently \cite{roland-th} a powerful Green's function formalism was published 
which unified earlier formulations \cite{mcmillan,ishii,furusaki1,furusaki} 
and improved upon them. 
First applications were made for NS, SNS, and SNSNS systems 
\cite{roland-th,miriam,roland2}. 
An important feature of these calculations was, that the systems were chosen 
to have a finite transverse width and they were focussed on particular 
phenomena which occur at specially chosen transverse widths. 
Up to now only treatments are known for an 
infinite width \cite{6:Tanaka91,stojkovic,miller}, or if a finite width 
was considered it was done in a global way, in terms of the number 
of allowed transverse modes \cite{furusaki,beenakker,lesovik}. 

\begin{figure}[htbp]
\centerline{\epsfig{figure=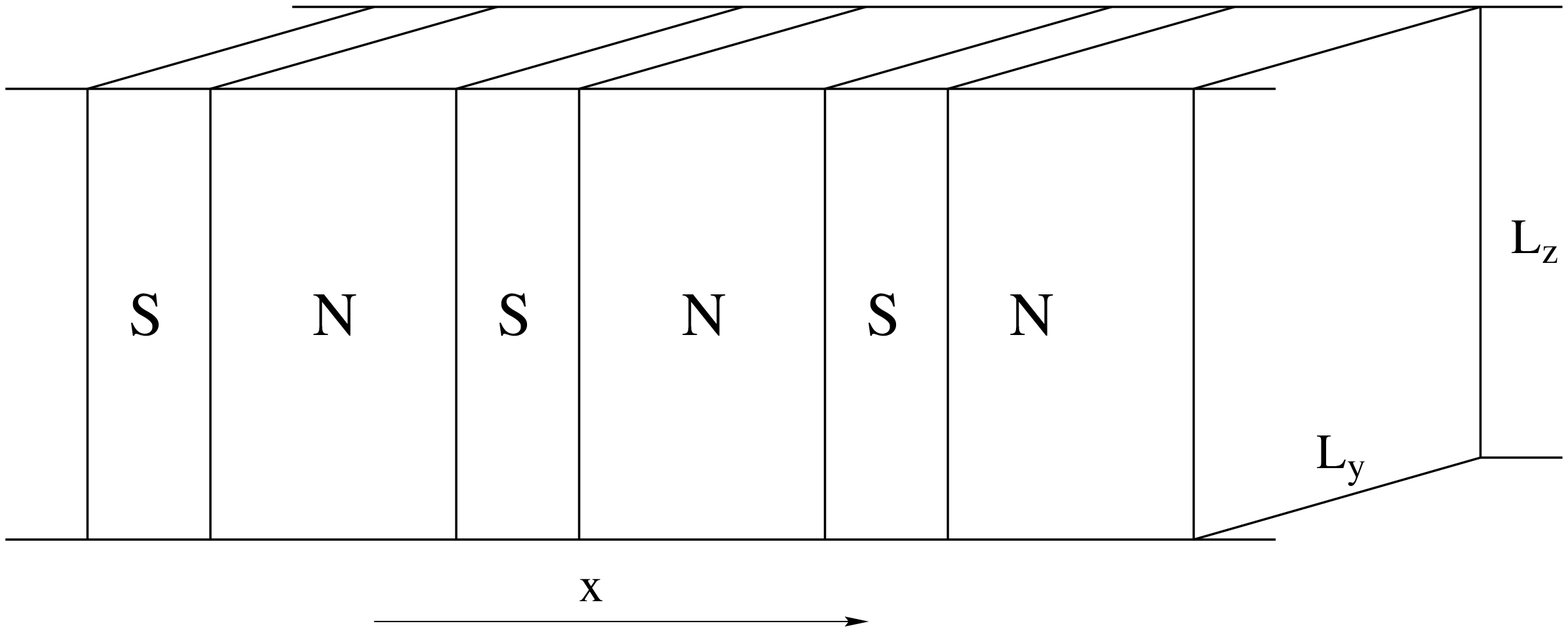,angle=0,width=9.0cm}}
\caption[]{Illustration of a Superconductor/Normal metal multilayer with 
finite transverse widths. } 
\label{MultilayerIntro}
\end{figure}

The aim of the present paper is to show applications to different SN 
multilayer structures, by which we understand a periodic sequence of S 
and N layers, extended in the $x$-direction, see Fig. \ref{MultilayerIntro}. 
A set of observables is calculated, such as the local density of 
states (LDOS) in the S and N layers of a multilayer, and 
the supercurrent in the multilayer. 

In Section \ref{s:theory}, we present the theory, applied to 
SN multilayers. 
In many situations, it is enough to work within the Andreev 
approximation (AA), 
which reduces to taking into account only Andreev reflections at the S/N 
interfaces. In Section \ref{s:andreev} we will show results derived 
within AA. Exact calculations, which include ordinary reflections, 
are discussed in Section \ref{s:exact}, in relation with the so-called 
critical transverse widths, at which a new transverse mode is starting 
to contribute. 
In Section \ref{s:currents} we present results for the supercurrent in 
the SN multilayers. 
In Section \ref{s:selfcons} we study the consequences of 
using a selfconsistently calculated gap. 
Finally, to complete the picture, in Section \ref{s:interface} we take into 
consideration an interface potential, by this modelling a Schottky barrier. 
We apply a simple $\delta$-function barrier located at the SN 
interface, introduced by Blonder {\it et al.} \cite{blonder}. 

\section{Clean SN structures}
\label{s:theory} 

The purpose of this section is to summarize first the basic ingredients of the 
general theory presented in Ref. \cite{roland-th} which we need in 
calculating the LDOS $\rho(x,E)$ for energies $E$ of the order of magnitude 
of the gap energy and the supercurrent $I$. 
After that we show how the general theory is elaborated for applications to 
SN multilayers. 

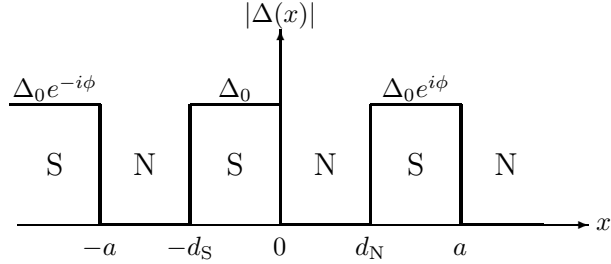
\begin{figure}[t]
   \begin{center}
    \setlength{\unitlength}{1mm}
    \begin{picture}(80,29)(-40,4)
      \put( 00,31){\makebox(0,0)[b]{$|\Delta(x)|$}}
      \put(-30,21){\makebox(0,0)[b]{$\Delta_0e^{-i\phi}$}}
      \put(-06,21){\makebox(0,0)[b]{$\Delta_0$}}
      \put( 18,21){\makebox(0,0)[b]{$\Delta_0e^{i\phi}$}}
      \multiput(-30,12)(24,0){3}{\makebox(0,0){\large S}}
      \multiput(-18,12)(24,0){3}{\makebox(0,0){\large N}}
      \thicklines
      \multiput(-36,20)(24,0){3}{\line(1,0){12}}
      \multiput(-24,04)(12,0){5}{\line(0,1){16}}
      \multiput(-24,04)(24,0){2}{\line(1,0){12}}
      \put(24,04){\line(1,0){11}}
      \thinlines
      \put( 00,04){\vector(0,1){26}}
      \put(-35,04){\line(1,0){70}}
      \put( 35,04){\vector(1,0){6}}
      \put( 42,04){\makebox(0,0)[l]{$x$}}
      \put(-24,00){\makebox(0,0)[b]{$-a_{\vphantom{\rm N}}$}}
      \put(-12,00){\makebox(0,0)[b]{$-d_{\rm S}$}}
      \put( 00,00){\makebox(0,0)[b]{$0_{\vphantom{\rm N}}$}}
      \put( 12,00){\makebox(0,0)[b]{$d_{\rm N}$}}
      \put( 24,00){\makebox(0,0)[b]{$a_{\vphantom{\rm N}}$}}
    \end{picture}
   \end{center}
   \caption{The Kronig-Penney model for the pair potential, used in
            Ref.\ \protect\cite{6:Tanaka91}.}
   \label{fg:tanaka}
  \end{figure}

We study a periodic SN multilayer, which extends in the $x$- direction, 
as depicted in Fig. \ref{MultilayerIntro}. 
In the transverse directions $y$ and $z$, the system has finite size, 
$L_y=L_z=L_t$. 
We apply a Kronig-Penney superlattice model, depicted in 
Fig. \ref{fg:tanaka}, which means that the pair potential is 
\begin{eqnarray}
\Delta(x+d_S+d_N)=\Delta(x)e^{i\phi} \nonumber \\
\Delta(x)=\left\{ \begin{array}{ll}
        \Delta & \mbox{if x $\in$ S layer}\\
        0      & \mbox{if x $\in$ N layer}, 
                  \end{array}
          \right.
\label{eq1}
\end{eqnarray}
where $d_S$ and $d_N$ are the thicknesses of the S and N layers respectively, 
and $\phi$ is the phase difference between two neighbouring 
superconducting layers. 

Both quantities we want to calculate, can be expressed 
in terms of the Green's function in the following way 
\begin{equation}\label{cs:dos}
\rho(x,E)=-\frac{2}{\pi}\frac{1}{L_yL_z}
\lim_{\delta\rightarrow 0}\sum_{k_y,k_z}
\mbox{Im}\,G_{11}(x,x;k_y,k_z,E+i\delta).
\end{equation}

\begin{equation}\label{cs:current}
I=-2ie\frac{1}{L_yL_z}\sum_{k_y,k_z}\lim_{x'\rightarrow x}
\left(\frac{\partial}{\partial x'}-\frac{\partial}{\partial x}\right)
kT\sum_n G_{11}(x,x';k_y,k_z,i\omega_n).
\end{equation}
where the Green's function $G_{11}$ is the upper left element of the 
matrix Green's function. 

The Green's function can be expressed in terms of a kind of 
wave function, which is the 
solution of the one-dimensional differential equation 
\begin{equation}\label{wavefct}
\left( \begin{array}{cc}
\displaystyle 
i\omega_n+\frac{d^2}{dx^2}+k_{Fx}^2 & -\Delta\\
-\Delta^{\ast} & \displaystyle i\omega_n-\frac{d^2}{dx^2}-k_{Fx}^2
\end{array} \right)\Psi(x)
=0.
\end{equation}
where $k_{Fx}^2\equiv\mu-k_y^2-k_z^2$. 
Note that the Bogoliubov equations arise from Eq. (\ref{wavefct}) 
by substituting $i\omega_n\rightarrow E$. 
The solution of Eq. (\ref{wavefct}) for the spatially dependent $\Delta$ 
of the multilayer (\ref{eq1}) is built up by starting with the solution 
for a homogeneous superconductor, having a constant $\Delta_S$. 
The wave function can then be written as 
\begin{equation}
\label{psi}
\Psi_{\rm S}^{\sigma\nu}(x)=\Biggl(\begin{array}{c}
u_{\rm S}^{\sigma}e^{i\phi_{\rm S}/2}\vspace{2mm}\\
u_{\rm S}^{-\sigma}e^{-i\phi_{\rm S}/2}
\end{array}\Biggr) e^{i\sigma\nu k_{\rm S}^{\sigma}x},
\end{equation}
with $u_{\rm S}^{\sigma}=\sqrt{i\omega_n+i\sigma\Omega_{\rm S}}$, 
$i\Omega_{S}=\sqrt{(i\omega_n+0^+)^2-\Delta_{S}^2}$, and the Matsubara 
frequencies $\omega_n=n\pi k_BT$, $n$ taking odd integer values only. 
The four standard solutions are labeled with the sign indices $\sigma$
and $\nu$, that can both equal $\pm 1$.
The index $\sigma$ refers to the type
of the propagating particle (electron-like for $\sigma=+$ and hole-like for
$\sigma=-$) and the index $\nu$ indicates the direction of
propagation.

In order to express the Green's function in terms of
$\Psi_{\rm S}^{\sigma\nu}(x)$, a conjugate wave function is needed, namely,
\begin{equation}
\label{psitilde}
\tilde{\Psi}_{\rm S}^{\sigma\nu}(x)=\Biggl(\begin{array}{cc}
~~u_{\rm S}^{\sigma}e^{-i\phi_{\rm S}/2}&
~~u_{\rm S}^{-\sigma}e^{i\phi_{\rm S}/2}
\end{array}\Biggr) e^{i\sigma\nu k_{\rm S}^{\sigma}x} 
\end{equation}
(which is not the {\em hermitian\/} conjugate), 
in which $\nu$ has now to be explained as
minus the direction of propagation.

With the use of these wave functions, we can express the Green's function 
for a homogeneous superconductor as 
\begin{equation}\label{GS3}
{\cal G}_{S}(x,x')=\sum_{\sigma}d_{S}^{\sigma}
\Psi_{S}^{\sigma\mu}(x)\tilde{\Psi}_{S}^{\sigma,-\mu}(x'), 
\end{equation}
with $\mu=\mbox{sgn}(x-x')$ and 
$d_S^{\sigma}=-\frac{1}{4\Omega_Sk_S^{\sigma}}$. 

\subsection{A single interface and more interfaces}

For a single interface, situated at the position $x_j$, the general form 
of the Green's function is 
\begin{equation}\label{Gsi1}
G_{\nu j\nu'j}(x,x')={\cal G}_{\nu j}(x,x')\delta_{\nu\nu'}+
\sum_{\sigma\sigma'}d_{\nu j}^{\sigma}d_{\nu'j}^{\sigma'}
\Psi_{\nu j}^{\sigma\nu}(x)
t_{\nu j\nu'j}^{\sigma\sigma'\nu\nu'}
\tilde{\Psi}_{\nu'j}^{\sigma'\nu'}(x'),
\end{equation}
where the subscript $(\nu j)$ refers to the part of the system 
that is at the $\nu$ side of the interface at position $x_j$. 
The first term accounts for the possible ways of propagating from $x'$
to $x$ without being scattered at the interface, while 
the second term describes the propagation via the interface. 

The scattering matrix $t_{\nu j\nu'j'}^{\sigma\sigma'\nu\nu'}$ is found
by applying boundary conditions at $x=x_j$. 
We require the continuity of the Green's function and its derivative. 
Thus, the equation obeyed by the $t$-matrix is 
 \begin{equation}
    \sum_{\sigma\nu}\nu d_{\nu j}^{\sigma}
      \Psi_{\nu j}^{\sigma\nu}(x_j)\,
      t_{\nu j\nu'j}^{\sigma\sigma'\nu\nu'}=
    -\nu'\Psi_{\nu'j}^{\sigma',-\nu'}(x_j). 
  \end{equation}

If we consider a system with an arbitrary number of interfaces, 
with position coordinates $x_j<x_{j+1}$, then the scattering of the 
quasiparticles is described by the scattering matrices 
$T_{\nu j\nu'j'}^{\sigma\sigma'\mu\mu'}$. The general form of the 
Green's function is 
\begin{eqnarray}&&
G_{\nu j\nu'j'}(x,x')={\cal G}_{\nu j}(x,x')\left[
\delta_{\nu\nu'}\delta_{jj'}+\delta_{-\nu\nu'}\delta_{j+\nu,j'}
\right]
\nonumber\vspace{2mm}\\\label{cs:Gmi1}&&\hspace{3cm}
\mbox{}+\sum_{\sigma\sigma'}\sum_{\mu\mu'}
d_{\nu j}^{\sigma}d_{\nu'j'}^{\sigma'}
\Psi_{\nu j}^{\sigma\mu}(x)
T_{\nu j\nu'j'}^{\sigma\sigma'\mu\mu'}
\tilde{\Psi}_{\nu'j'}^{\sigma'\mu'}(x').
\end{eqnarray}

Again, imposing the boundary conditions, we obtain a Lippmann--Schwinger 
equation which allows to calculate the $T$-matrices by means of the 
single interface $t$-matrices 
\begin{equation}\label{Tmi1}
T_{\nu j\nu'j'}^{\sigma\sigma'\nu\mu'}=
t_{\nu j\mu'j}^{\sigma\sigma'\nu\mu'}\left[
\delta_{\mu'\nu'}\delta_{jj'}+\delta_{-\mu'\nu'}\delta_{j+\mu',j'}
\right]+\sum_{\sigma''\nu''}
t_{\nu j\nu''j}^{\sigma\sigma''\nu\nu''}
d_{\nu''j}^{\sigma''}
T_{\nu''j\nu'j'}^{\sigma''\sigma',-\nu''\mu'}.
\end{equation}

This equation expresses the idea of multiple scattering, since 
the matrix $T_{\nu j\nu'j'}^{\sigma\sigma'\nu\mu'}$ 
contains all possible processes that yield the correct final state.
The first term in Eq.\ (\ref{Tmi1}) accounts for the possibility that
the particle is scattered once.
The second term collects the processes in which the particle is scattered
once due to $t_{\nu j\nu''j}^{\sigma\sigma''\nu\nu''}$ and an
arbitrary number of other times due to
$T_{\nu''j\nu'j'}^{\sigma''\sigma',-\nu''\mu'}$.

\subsection{Periodic SN multilayers}
\label{s:multilayers}

Till now we just summarized the description given previously \cite{roland-th}. 
We will now focus on an infinite periodic SN multilayer to which 
the theory has not been applied yet. 
For an infinite multilayer, it is always possible to refer to any layer 
of the system by referring to an even-numbered (or an odd-numbered 
interface only). This can lead to a further simplification of 
the Lippmann--Schwinger equation. 
We choose to refer to any part of the system by referring 
the even interfaces. 
However, for the $t$-matrices, both even and odd interface indices 
need to be used. 

In order to rewrite Eq. (\ref{Tmi1}) for the present purpose, 
we define the following matrices 
\begin{eqnarray}
& & {\mathbf T}_{jj'}\equiv{\mathbf D}_j\delta_{jj'}+
    \left(\begin{array}{cc}
      T_{\mu j\mu'j'}^{\sigma\sigma',-\mu\mu'}&
      T_{\mu j,-\mu'j'}^{\sigma\sigma',-\mu\mu'}\vspace{2mm}\\
      T_{-\mu j\mu'j'}^{\sigma\sigma',-\mu\mu'}&
      T_{-\mu j,-\mu'j'}^{\sigma\sigma',-\mu\mu'}
    \end{array}\right)
  \label{def:TDABC}\\
& & {\mathbf D}_j\equiv
    \left(\begin{array}{cc}
      \frac{\displaystyle\delta_{\sigma\sigma'}\delta_{\mu\mu'}}
           {\displaystyle d_{\mu j}^{\sigma}}&0\\0&
      \frac{\displaystyle\delta_{\sigma\sigma'}\delta_{\mu\mu'}}
           {\displaystyle d_{-\mu j}^{\sigma}}
    \end{array}\right)
  \\
& & {\mathbf A}_j\equiv
    \Biggl(\begin{array}{cc}~~0&
      ~~t_{-\mu,j+\mu,\mu',j+\mu}^{\sigma\sigma',-\mu\mu'}
        d_{-\mu',j-2}^{\sigma'}\delta_{\mu-}\delta_{\mu'-}\\~~0&~~0
    \end{array}\Biggr)
  \\
& & {\mathbf B}_j\equiv
    \Biggl(\begin{array}{cc}~~0&
      ~~t_{-\mu,j+\mu,\mu',j+\mu}^{\sigma\sigma',-\mu\mu'}
        d_{-\mu'j}^{\sigma'}\delta_{-\mu\mu'}\\
      ~~t_{-\mu j\mu'j}^{\sigma\sigma',-\mu\mu'}d_{\mu'j}^{\sigma'}&~~0
    \end{array}\Biggr)
  \\
& & {\mathbf C}_j\equiv
    \Biggl(\begin{array}{cc}~~0&
      ~~t_{-\mu,j+\mu,\mu',j+\mu}^{\sigma\sigma',-\mu\mu'}
        d_{-\mu',j+2}^{\sigma'}\delta_{\mu+}\delta_{\mu'+}\\~~0&~~0
    \end{array}\Biggr).
  \end{eqnarray}

All elements of these $2\times 2$ matrices can themselves be
regarded as $4\times 4$ matrices, with the indices $(\sigma,\mu)$ and
$(\sigma',\mu')$ labeling the rows and the columns, respectively.
That makes ${\mathbf T}_{jj'}$ an $8\times 8$ matrix that satisfies
  \begin{equation}\label{cs:Tml3}
    {\mathbf A}_j\cdot{\mathbf T}_{j-2,j'}+({\mathbf B}_j-{\mathbf 1})\cdot{\mathbf T}_{jj'}+
    {\mathbf C}_j\cdot{\mathbf T}_{j+2,j'}+{\mathbf D}_j\delta_{jj'}=0.
  \end{equation}
In terms of these matrices, the system Green's function (\ref{cs:Gmi1}) 
gets the following form 
\begin{eqnarray}&&
G_{\nu j\nu'j'}(x,x')={\cal G}_{\nu j}(x,x')\left[
\delta_{\nu\nu'}\delta_{jj'}+\delta_{-\nu\nu'}\delta_{j+\nu,j'}
\right]
\nonumber\vspace{2mm}\\
&&\hspace{0cm}
\mbox{}+\sum_{\sigma\sigma'}\sum_{\mu\mu'}
d_{\nu j}^{\sigma}d_{\nu'j'}^{\sigma'}
\Psi_{\nu j}^{\sigma\mu}(x)
\nonumber\vspace{2mm}\\\label{cs:Gmi2}&&\hspace{.5cm}
({\mathbf T}_{jj'}-{\mathbf D}_j\delta_{jj'})_{\nu j\nu'j'}^{\sigma\sigma'\mu\mu'}
(...A_{j},B_{j},C_{j},A_{j+1},B_{j+1},C_{j+1},...)
\tilde{\Psi}_{\nu'j'}^{\sigma'\mu'}(x'), 
\end{eqnarray} 

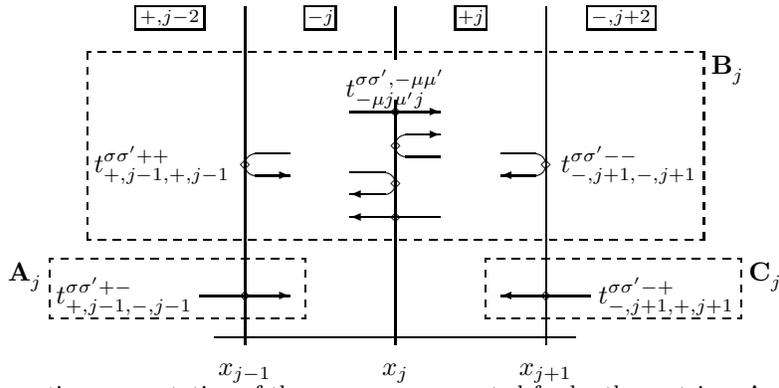
\begin{figure}[t]
   \begin{center}
    \setlength{\unitlength}{1mm}
    \begin{picture}(80,44)(-40,4)
      \put(-34.5,46){\framebox(9,3){$\scriptstyle+,j-2$}}
      \put(-12,46){\framebox(4,3){$\scriptstyle-j$}}
      \put( 08,46){\framebox(4,3){$\scriptstyle+j$}}
      \put( 25.5,46){\framebox(9,3){$\scriptstyle-,j+2$}}
      \put( 00,42){\line(0,1){7}}
      \put(-20,18){\begin{picture}(40,25)(-20,00)
        \put( 42,  25  ){\makebox(0,0)[tl]{${\mathbf B}_j^{\vphantom{X}}$}}
        \put( 18.5,08.5){\vector(-1,0){4.5}}
        \put( 18.5,10  ){\oval(3,3)[r]}
        \put( 20,  10  ){\makebox(0,0){$\scriptstyle\diamond$}}
        \put( 22,  10  ){\makebox(0,0)[l]{$t_{-,j+1,-,j+1}^{\sigma\sigma'--}$}}
        \put( 18.5,11.5){\line(-1,0){4.5}}
        \put(-18.5,11.5){\line(1,0){4.5}}
        \put(-22,  10  ){\makebox(0,0)[r]{$t_{+,j-1,+,j-1}^{\sigma\sigma'++}$}}
        \put(-20,  10  ){\makebox(0,0){$\scriptstyle\diamond$}}
        \put(-18.5,10  ){\oval(3,3)[l]}
        \put(-18.5,08.5){\vector(1,0){4.5}}
        \put( 00,  19  ){\makebox(0,0)[b]
                        {$t_{-\mu j\mu'j}^{\sigma\sigma',-\mu\mu'}$}}
        \put( 00,  17  ){\line(-1,0){6}}
        \put( 00,  17  ){\makebox(0,0){$\scriptstyle\diamond$}}
        \put( 00,  17  ){\vector(1,0){6}}
        \put( 01.5,14  ){\vector(1,0){4.5}}
        \put( 01.5,12.5){\oval(3,3)[l]}
        \put( 00,  12.5){\makebox(0,0){$\scriptstyle\diamond$}}
        \put( 01.5,11  ){\line(1,0){4.5}}
        \put(-01.5,09  ){\line(-1,0){4.5}}
        \put(-01.5,07.5){\oval(3,3)[r]}
        \put( 00,  07.5){\makebox(0,0){$\scriptstyle\diamond$}}
        \put(-01.5,06  ){\vector(-1,0){4.5}}
        \put( 00,  03  ){\vector(-1,0){6}}
        \put( 00,  03  ){\makebox(0,0){$\scriptstyle\diamond$}}
        \put( 00,  03  ){\line(1,0){6}}
        \put( 00,  00  ){\line(0,1){18.5}}
        \put(-41,  00  ){\dashbox{1}(82,25)[tr]{}}
      \end{picture}}
      \put(-47,15.5){\makebox(0,0)[tr]{${\mathbf A}_j^{\vphantom{X}}$}}
      \put( 47,15.5){\makebox(0,0)[tl]{${\mathbf C}_j^{\vphantom{X}}$}}
      \put(-27,10.5){\makebox(0,0)[r]{$t_{+,j-1,-,j-1}^{\sigma\sigma'+-}$}}
      \put(-20,10.5){\vector(1,0){6}}
      \put(-20,10.5){\makebox(0,0){$\scriptstyle\diamond$}}
      \put(-20,10.5){\line(-1,0){6}}
      \put( 20,10.5){\line(1,0){6}}
      \put( 20,10.5){\makebox(0,0){$\scriptstyle\diamond$}}
      \put( 20,10.5){\vector(-1,0){6}}
      \put( 27,10.5){\makebox(0,0)[l]{$t_{-,j+1,+,j+1}^{\sigma\sigma'-+}$}}
      \put(-46,07.5){\dashbox{1}(34,8)[tr]{}}
      \put( 12,07.5){\dashbox{1}(34,8)[tr]{}}
      \put(-24,05){\line(1,0){48}}
      \put(-20,04){\line(0,1){45}}
      \put( 00,04){\line(0,1){14}}
      \put( 20,04){\line(0,1){45}}
      \put(-20,00){\makebox(0,0)[b]{$x_{j-1}$}}
      \put( 00,00){\makebox(0,0)[b]{$x_j$}}
      \put( 20,00){\makebox(0,0)[b]{$x_{j+1}$}}
    \end{picture}
   \end{center}
   \caption{Schematic representation of the processes accounted for by
            the matrices ${\mathbf A}_j$, ${\mathbf B}_j$ and ${\mathbf C}_j$.}
   \label{fg:ABC}
  \end{figure}

All possible scattering processes are incorporated in Eq. (\ref{cs:Tml3}), 
which expresses the content of the Lippmann--Schwinger equation. 
To illustrate the different processes accounted by the ${\mathbf A}_j$, 
${\mathbf B}_j$, and ${\mathbf C}_j$ matrices, we show a schema in 
Fig. \ref{fg:ABC}. 

We now turn to the periodic system. 
For the moment, we assume that the phase of the pair potential is 
the same and equal to zero in all the S-layers. 
The periodicity 
allows us to simplify the problem and to rewrite equation (\ref{cs:Tml3}). 
First we perform the following transformations  
\begin{equation}\label{def:Tsi}
\hat{t}_{\nu j\nu'j}^{\sigma\sigma'\nu\nu'}\equiv
e^{i\sigma\nu k_{\nu j}^{\sigma}x_j}
t_{\nu j\nu'j}^{\sigma\sigma'\nu\nu'}
e^{i\sigma'\nu' k_{\nu'j}^{\sigma'}x_j}, 
\end{equation}
  \begin{equation}\label{def:Tmi}
      \hat{T}_{\mu j\mu'j'}^{\sigma\sigma'\nu\nu'}\equiv
      e^{i\sigma\nu k_{\mu j}^{\sigma}x_j}
      T_{\mu j\mu'j'}^{\sigma\sigma'\nu\nu'}
      e^{i\sigma'\nu' k_{\mu'j'}^{\sigma'}x_j'}. 
  \end{equation}
The scattering matrices with hats no longer depend on the interface 
positions $x_j$, although they still refer to the interface number, 
through the labels $\nu j$. 

The matrices ${\mathbf T}_{jj'}$, ${\mathbf D}_j$, ${\mathbf A}_j$, 
${\mathbf B}_j$, ${\mathbf C}_j$, become 
  \begin{eqnarray}
&&\hspace{-4mm}
    \hat{\mathbf T}_{jj'}\equiv
    \hat{\mathbf D}_j\delta_{jj'}+
    \left(\begin{array}{cc}
      \hat{T}_{\mu j\mu'j'}^{\sigma\sigma',-\mu\mu'}&
      \hat{T}_{\mu j,-\mu'j'}^{\sigma\sigma',-\mu\mu'}\vspace{2mm}\\
      \hat{T}_{-\mu j\mu'j'}^{\sigma\sigma',-\mu\mu'}&
      \hat{T}_{-\mu j,-\mu'j'}^{\sigma\sigma',-\mu\mu'}
    \end{array}\right)
  \label{def:TDABC2}\\
&&\hspace{-4mm}
    \hat{\mathbf D}\equiv
    \left(\begin{array}{cc}
      \frac{\displaystyle\delta_{\sigma\sigma'}\delta_{\mu\mu'}}
           {\displaystyle d_{\mu}^{\sigma}}
        e^{-i\sigma k_{\mu}^{\sigma}a_{\mu}}&0\\0&
      \frac{\displaystyle\delta_{\sigma\sigma'}\delta_{\mu\mu'}}
           {\displaystyle d_{-\mu}^{\sigma}}
        e^{-i\sigma k_{-\mu}^{\sigma}a_{-\mu}}
    \end{array}\right)
  \\&&\hspace{-4mm}
    \hat{\mathbf A}\equiv
    \Biggl(\begin{array}{cc}~~0&
      ~~\hat{t}_{\mu,-\mu'}^{\sigma\sigma',-\mu\mu'}
        d_{-\mu'}^{\sigma'}\delta_{\mu-}\delta_{\mu'-}
        e^{i\sigma'k_{-\mu'}^{\sigma'}a_{-\mu'}}\\~~0&~~0
    \end{array}\Biggr)
  \\&&\hspace{-4mm}
    \hat{\mathbf B}\equiv
    \Biggl(\begin{array}{cc}\hspace{-1cm}~~0&\hspace{-1cm}
      ~~\hat{t}_{\mu,-\mu'}^{\sigma\sigma',-\mu\mu'}
        d_{-\mu'}^{\sigma'}\delta_{-\mu\mu'}
        e^{i\sigma'k_{-\mu'}^{\sigma'}a_{-\mu'}}\\
      ~~\hat{t}_{-\mu\mu'}^{\sigma\sigma',-\mu\mu'}d_{\mu'}^{\sigma'}
        e^{i\sigma'k_{\mu'}^{\sigma'}a_{\mu'}}&\hspace{-1cm}~~0
    \end{array}\Biggr)
  \\&&\hspace{-4mm}
    \hat{\mathbf C}\equiv
    \Biggl(\begin{array}{cc}~~0&
      ~~\hat{t}_{\mu,-\mu'}^{\sigma\sigma',-\mu\mu'}
        d_{-\mu'}^{\sigma'}\delta_{\mu+}\delta_{\mu'+}
        e^{i\sigma'k_{-\mu'}^{\sigma'}a_{-\mu'}}\\~~0&~~0
    \end{array}\Biggr). 
  \end{eqnarray}

After these transformations, and as a consequence of the periodicity, 
the Green's function 
becomes dependent on the relative coordinates only, and 
the $j$-independent layer thicknesses 
$a_{\mu}=\mu(x_{j+\mu}-x_j)$ and $a_{-\mu}=\mu(x_j-x_{j-\mu})$ 
can be defined. 
In terms of these matrices, Eq.\ (\ref{cs:Tml3}) now reads as 
  \begin{equation}\label{cs:Tml4}
    \hat{\mathbf A}\cdot\hat{\mathbf T}_{j-2,j'}+
    (\hat{\mathbf B}-{\mathbf 1})\cdot\hat{\mathbf T}_{jj'}+
    \hat{\mathbf C}\cdot\hat{\mathbf T}_{j+2,j'}+
    \hat{\mathbf D}\delta_{jj'}=0.
  \end{equation}
This is a kind of discretized version of the original Green's function
equation.
The general solution is
  \begin{equation}\label{cs:Tml5}
    \hat{\mathbf T}_{jj'}=
    \hat{\mathbf X}_{{\rm sgn}(j-j')}^{|j-j'|/2}\cdot\hat{\mathbf T}_0,
  \end{equation}
where $\hat{\mathbf X}_{-}$, $\hat{\mathbf X}_{+}$ and 
$\hat{\mathbf T}_0$ are
implicitly given by the following set of equations
  \begin{eqnarray}
\label{def:Xminus}
&&    \hat{\mathbf A}\cdot\hat{\mathbf X}_{-}^2+
    (\hat{\mathbf B}-{\mathbf 1})\cdot\hat{\mathbf X}_{-}
+\hat{\mathbf C}=0
  \label{def:XXT}\\
\label{def:Xplus}
&&    \hat{\mathbf A}+(\hat{\mathbf B}-{\mathbf 1})
\cdot\hat{\mathbf X}_{+}+
    \hat{\mathbf C}\cdot\hat{\mathbf X}_{+}^2=0
  \\
\label{def:Tjj}
&&    [{\mathbf A}\cdot\hat{\mathbf X}_{-}+(\hat{\mathbf B}
-{\mathbf 1})+
    \hat{\mathbf C}\cdot\hat{\mathbf X}_{+}]\cdot
    \hat{\mathbf T}_0+\hat{\mathbf D}=0.
  \end{eqnarray}
Note that (\ref{def:Xminus}) and (\ref{def:Xplus}) are quadratic equations. 
By that, for a periodic system, we can rewrite the expression of the Green's 
function (\ref{cs:Gmi2}) in a simpler form 
\begin{eqnarray}&&
G_{\nu j\nu'j'}(x,x')={\cal G}_{\nu j}(x,x')\left[
\delta_{\nu\nu'}\delta_{jj'}+\delta_{-\nu\nu'}\delta_{j+\nu,j'}
\right]
\mbox{}+\sum_{\sigma\sigma'}\sum_{\mu\mu'}
d_{\nu j}^{\sigma}d_{\nu'j'}^{\sigma'}
\Psi_{\nu j}^{\sigma\mu}(x-x_j)
\nonumber\vspace{2mm}\\\label{cs:Gmi3}&&\hspace{1.5cm}
(\hat{\mathbf T}_{jj'}-
\hat{\mathbf D}_j\delta_{jj'})_{\nu j\nu'j'}^{\sigma\sigma'\mu\mu'}
(\hat{\mathbf A},\hat{\mathbf B},\hat{\mathbf C})
\tilde{\Psi}_{\nu'j'}^{\sigma'\mu'}(x'-x_{j'}), 
\end{eqnarray} 
The problem of calculating the LDOS or the supercurrent, 
reduces to solving the system of quadratic matrix equations 
(\ref{def:Xminus}) to (\ref{def:Tjj}). 

After some manipulations, one can manage to reduce the problem 
to solving a $2\times2$ matrix equation, which is equivalent to 
solving a system of 8 simultaneous equations with real coefficients. 
In the Appendix, we show this more explicitly. 

The formalism described up to now can be extended to the situation 
in which the pair potential has a phase, which allows for currents 
in the multilayer. 

By convention, we assume that the N-layer has also a phase, equal 
to the phase of one of the two adjacent S-layers. 
This means that the phase over a bilayer is constant and it makes a jump 
of $\phi$ at the edges between two bilayers. 

Suppose the interface between two bilayers in chosen at $j$ odd, then 
for even $j$ the $t$-matrices obey the equation 
 \begin{equation}\label{cs:ccm2}
    \sum_{\sigma\nu}\nu d_{\nu j}^{\sigma}
      u_{\nu j}^{\sigma\nu}\,
      \hat{t}_{\nu j\nu'j}^{\sigma\sigma'\nu\nu'}=
    -\nu'u_{\nu'j}^{\sigma',-\nu'}\hspace{6mm}(j~\mbox{even}).
  \end{equation}
while for odd interfaces, a $\phi$ dependence remains and we are left with
  \begin{equation}\label{cs:ccm3}
    \sum_{\sigma\nu}\nu d_{\nu j}^{\sigma}
      U(\nu\phi\delta_{-\nu\nu'})\,
      u_{\nu j}^{\sigma\nu}\,
      \hat{t}_{\nu j\nu'j}^{\sigma\sigma'\nu\nu'}=
    -\nu'u_{\nu'j}^{\sigma',-\nu'}\hspace{6mm}(j~\mbox{odd}),
  \end{equation}
where 
 \begin{equation}\label{def:Uu}
    U(\phi)\equiv\left(\begin{array}{cccc}
      ~~e^{i\phi/2}&
      ~~\hspace{-5mm}0&
      ~~\hspace{-5mm}0&
      ~~\hspace{-5mm}0\vspace{2mm}\\
      ~~0&
      ~~\hspace{-5mm}e^{-i\phi/2}&
      ~~\hspace{-5mm}0&
      ~~\hspace{-5mm}0\vspace{2mm}\\
      ~~0&
      ~~\hspace{-5mm}0&
      ~~\hspace{-5mm}e^{i\phi/2}&
      ~~\hspace{-5mm}0\vspace{2mm}\\
      ~~0&
      ~~\hspace{-5mm}0&
      ~~\hspace{-5mm}0&
      ~~\hspace{-5mm}e^{-i\phi/2}
    \end{array}\right).
 \end{equation}

\section{Local density of states in SN multilayers}
\label{s:andreev}

First we apply the theory in calculating 
the LDOS in the middle of one of the S or N layers. 
For most of the systems which will be described, the transverse width 
$L_t$ is fixed to 13 Bohr, the chemical potential $\mu=0.5$ Ryd, and the 
LDOS is normalized to the spatially constant LDOS of the bulk N material. 
The coupling potential V is calculated using the BCS formula 
\begin{equation}
T_c=1.13\omega_{D}e^{-1/N(\mu)V}, 
\end{equation}
where $N(\mu)=\sqrt{\mu}/4\pi$. 
For Al with $T_c=1.2$ K, $\mu=0.5$ Ryd, and $\omega_D=375$ K, 
we find $V=9.516$ Ryd. For the pair potential $\Delta$ we choose 
a value of 0.0001 Ryd. Given the BCS relation $\frac{\Delta}{k_BT_c}=1.77$, 
the pair function should be somewhat larger, $\Delta=0.00018$ Ryd. 
However, we should also keep in mind the reduction of 
the pair function due to the finite size of the system. 
This we will discuss in Section \ref{s:selfcons}, in which we will 
determine the gap function selfconsistently. 
Furthermore, many of the results we will show are not quite sensitive to 
the precise choice for $\Delta$. 

It appears that the LDOS curves for SN multilayers look rather complicated. 
As a preparation to understand them, we first look at simpler systems and we 
postpone the treatment of SN multilayers to subsections \ref{subsmultil} 
and \ref{subsinf}. 
In the coming subsection, we will follow the development of the 
LDOS for a bulk system to the LDOS for systems with a few interfaces. 

\subsection{From bulk superconductor to SNS system}
\label{subssns}

In this subsection we will first follow the development of the LDOS for 
a bulk bar-shaped superconductor to the LDOS of a SN multilayer with 
$d_S>>(\xi,d_N$) and of multilayers with $d_S>>\xi$, in which $d_N$ becomes 
comparable to $d_S$. 
For the present clean systems the BCS coherence length $\xi\approx 4000$ 
Bohr. Looking at Fig. \ref{fg:tanaka} it is clear that a multilayer with thick 
superconducting layers, such that $d_S>>\xi$, comes close to a SNS system, 
particularly for $E<\Delta$. 

\begin{figure}[htbp]
\centerline{\epsfig{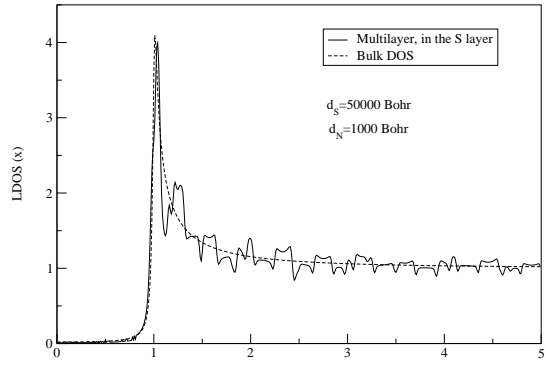}}
\caption[]{LDOS for a SN multilayer ($d_S=50000$ Bohr and $d_N=1000$ Bohr) 
and for a bar shaped superconductor (dashed line).}
\label{BulkS}
\end{figure}

In Fig. \ref{BulkS} we show the LDOS inside the S layer 
of a SN multilayer and the DOS of bulk S material. 
One clearly sees the singularity in the LDOS at $E=\Delta$. 
The non-zero DOS of the S material for $E<\Delta$ is due to the small 
imaginary part $i\delta$ added to the energy, $E+i\delta$. 
In all calculations we used $\delta\approx 0.02\Delta$. 
Here $d_N=1000$ Bohr and $d_S=50000$ Bohr, so that the presence 
of the N layer is just a small 
perturbation from a bar shaped superconductor. 

\begin{figure}[htbp]
\centerline{\epsfig{figure=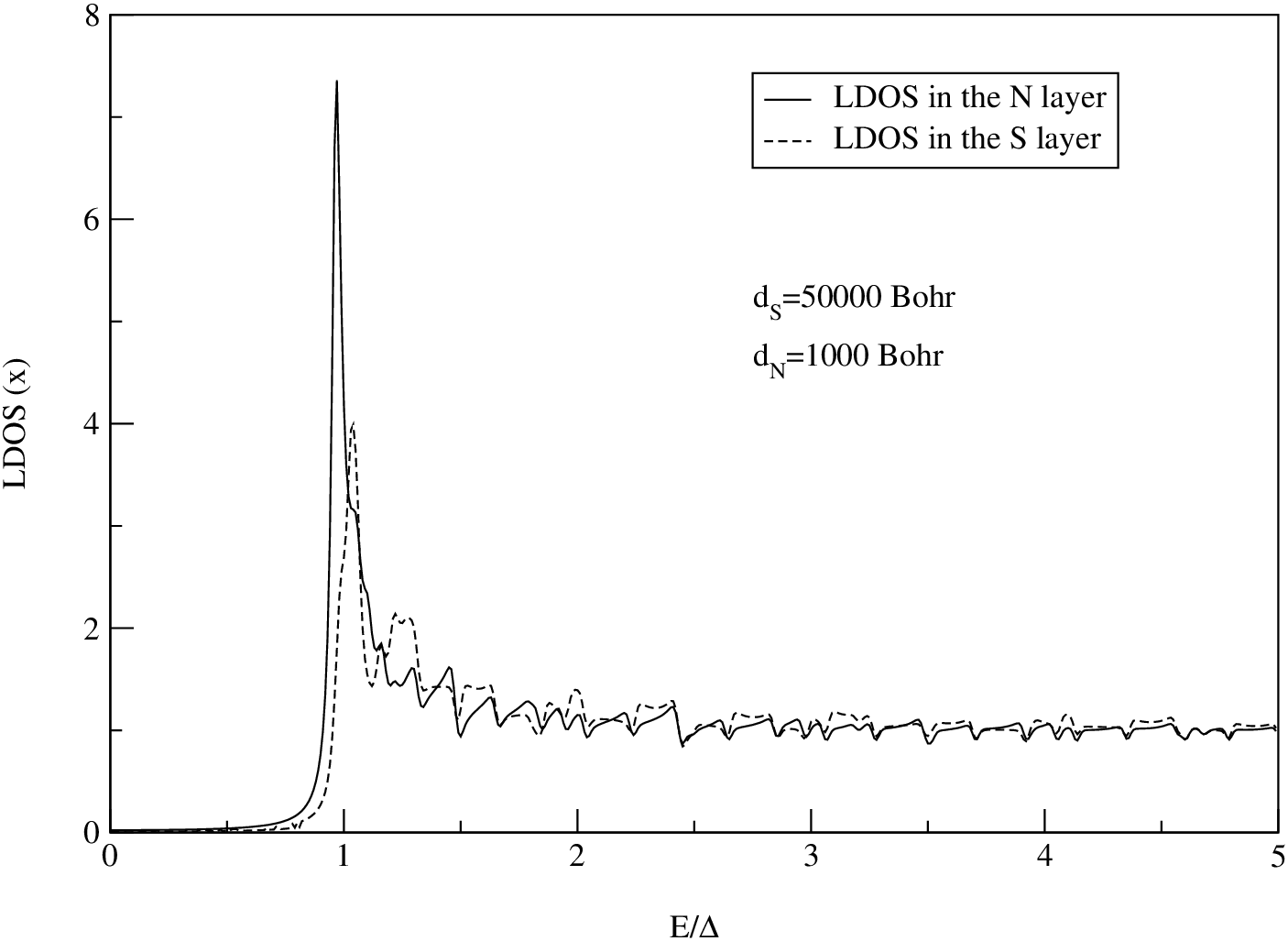,angle=0,width=7.1cm}}
\caption[]{LDOS for a SN multilayer ($d_S=50000$ Bohr and $d_N=1000$ Bohr) 
in the N layer (solid line) and S layer (dashed line). }
\label{dN1dS50}
\end{figure}

First we concentrate on the development of the LDOS for $E<\Delta$. 
In Fig. \ref{dN1dS50} we show for the same system both the LDOS 
inside the S and the N layer. Due to the very small N layer thickness, 
there is just one Andreev bound state in the N layer LDOS close to the 
gap value of the energy, which is broadened by $\delta$ to a peak. 

\begin{figure}[htbp]
\centerline{\epsfig{figure=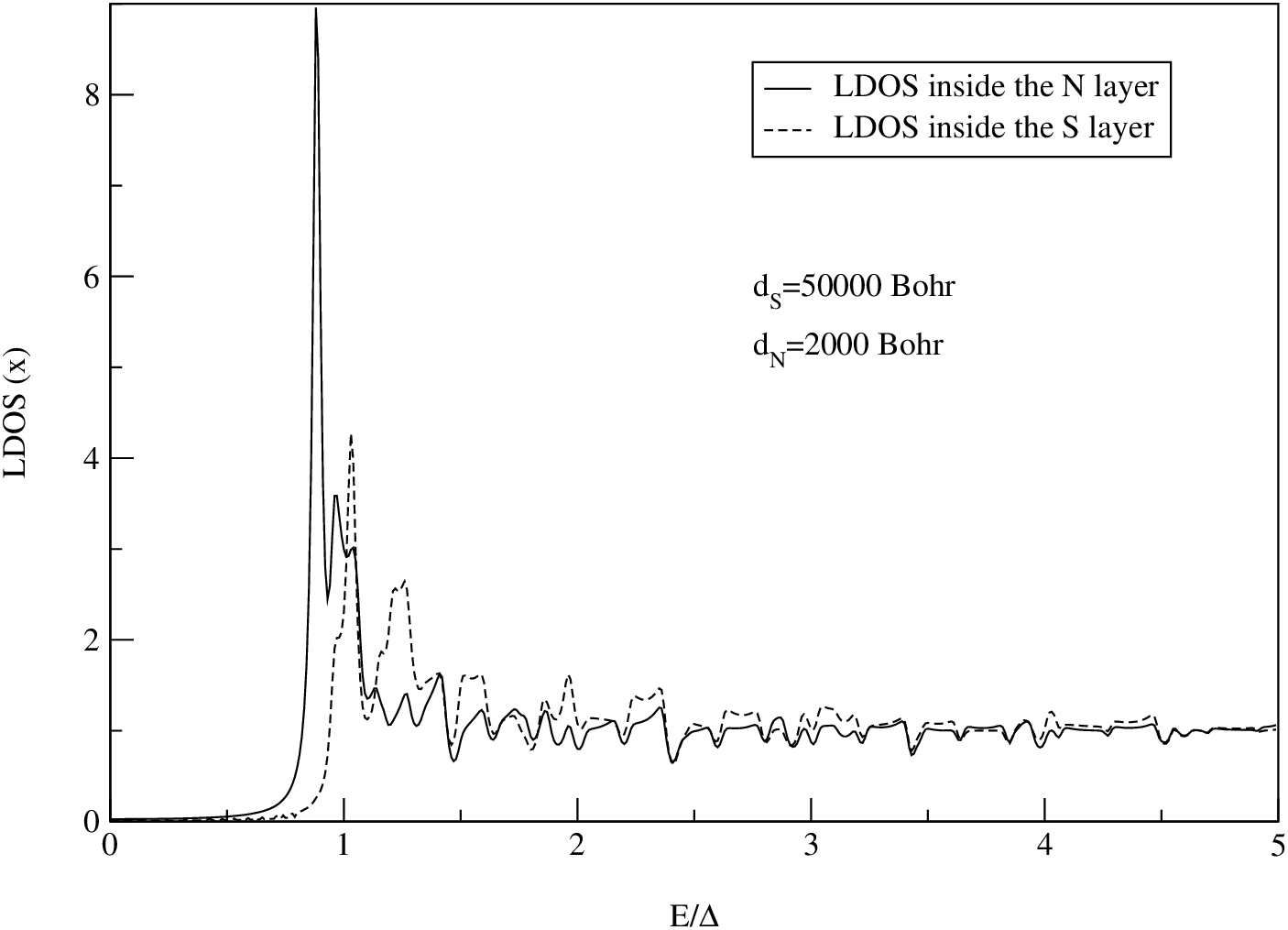,angle=0,width=7.1cm}}
\caption[]{LDOS for a SN multilayer ($d_S=50000$ Bohr and $d_N=2000$ Bohr) 
in the N layer (solid line) and S layer (dashed line). }
\label{dN2dS50}
\end{figure}

\begin{figure}[htbp]
\centerline{\epsfig{figure=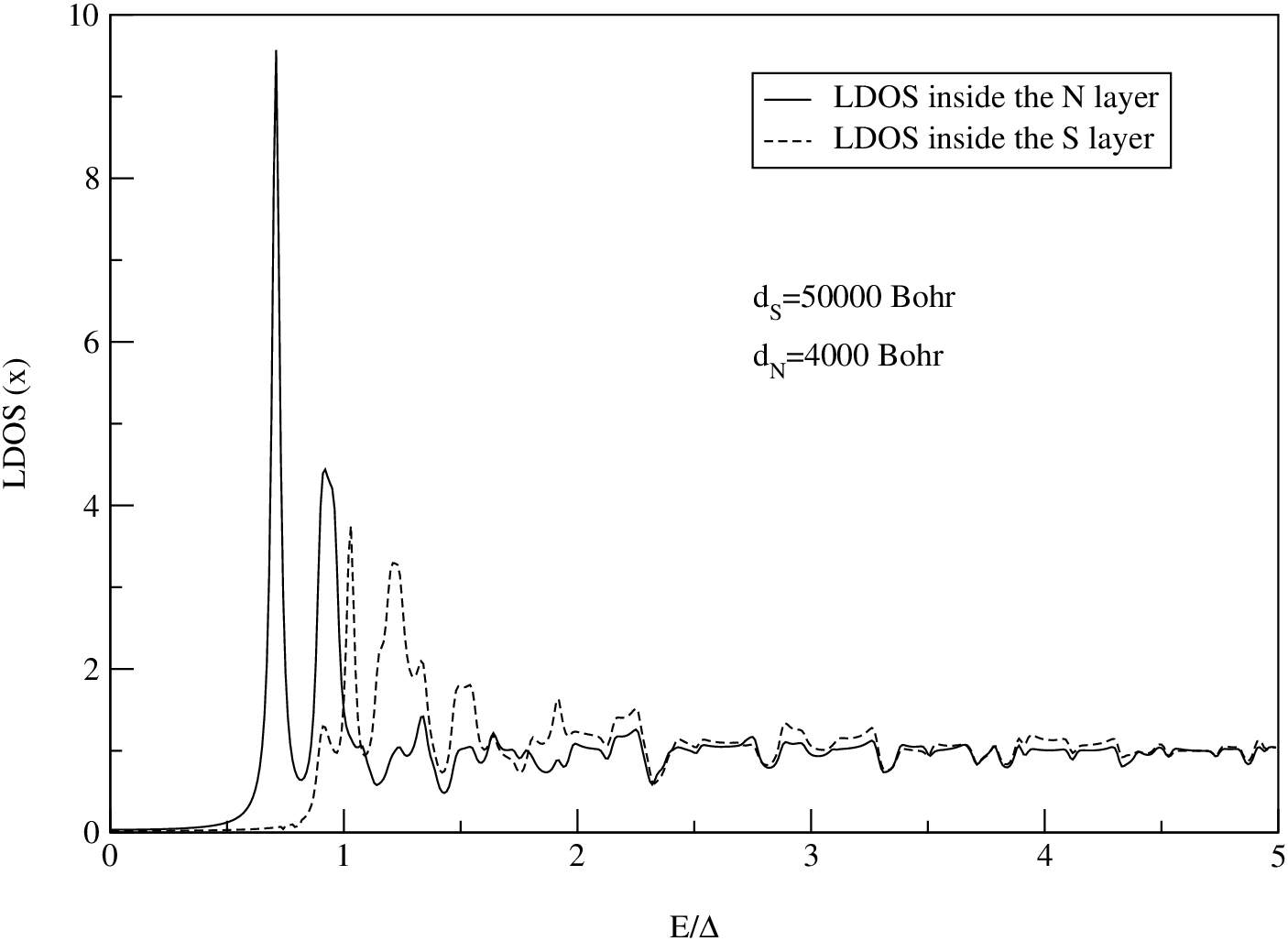,angle=0,width=7.1cm}}
\caption[]{LDOS for a SN multilayer ($d_S=50000$ Bohr and $d_N=4000$ Bohr) 
in the N layer (solid line) and S layer (dashed line). }
\label{dN4dS50}
\end{figure}

\begin{figure}[htbp]
\centerline{\epsfig{figure=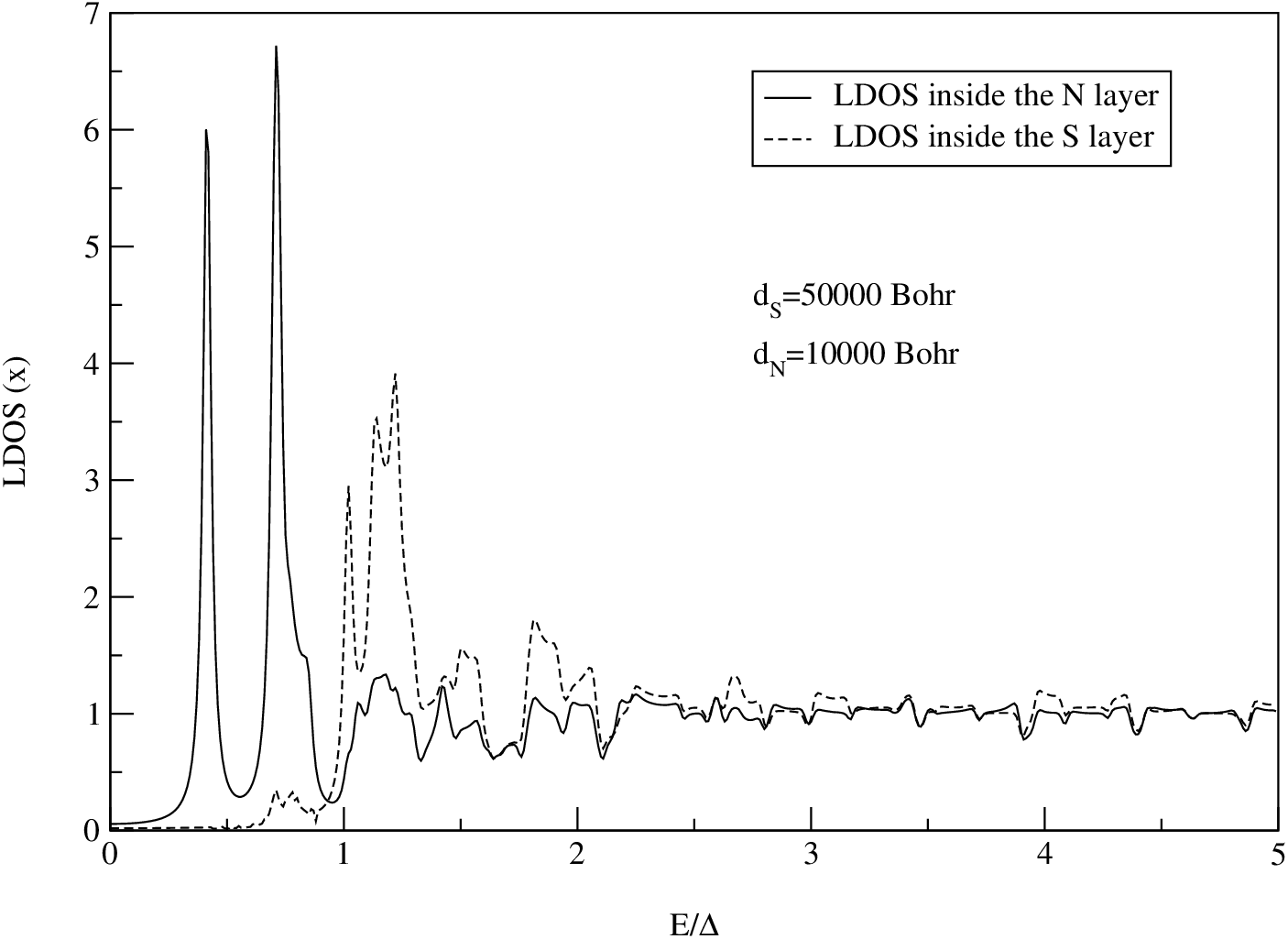,angle=0,width=7.1cm}}
\caption[]{LDOS for a SN multilayer ($d_S=50000$ Bohr and $d_N=10000$ Bohr) 
in the N layer (solid line) and S layer (dashed line). }
\label{dN10dS50}
\end{figure}

Figs. \ref{dN2dS50}, \ref{dN4dS50}, and \ref{dN10dS50} show what happens to 
the LDOS of the SN multilayers if we increase $d_N$ to 
$d_{N}=$ 2000, 4000, and 10000 Bohr respectively. 
The pictures look more and more complex as we increase $d_N$. 
In the N layer LDOS we notice the appearance of more Andreev bound 
states at lower and lower energies. 
The singularity in the S-layer DOS lowers, certainly due to reduced 
interaction of the neighbouring S-layers. 

The oscillations in the LDOS in all figures for $E>\Delta$ are a 
periodic-multilayer effect. 
Before discussing this band structure effect, we will 
first consider multilayer systems with decreasing $d_S$, 
by that making more explicit the multilayer character of the system. 

\subsection{From SNS system to SN multilayer} 
\label{subsmultil}

\begin{figure}[htbp]
\centerline{\epsfig{figure=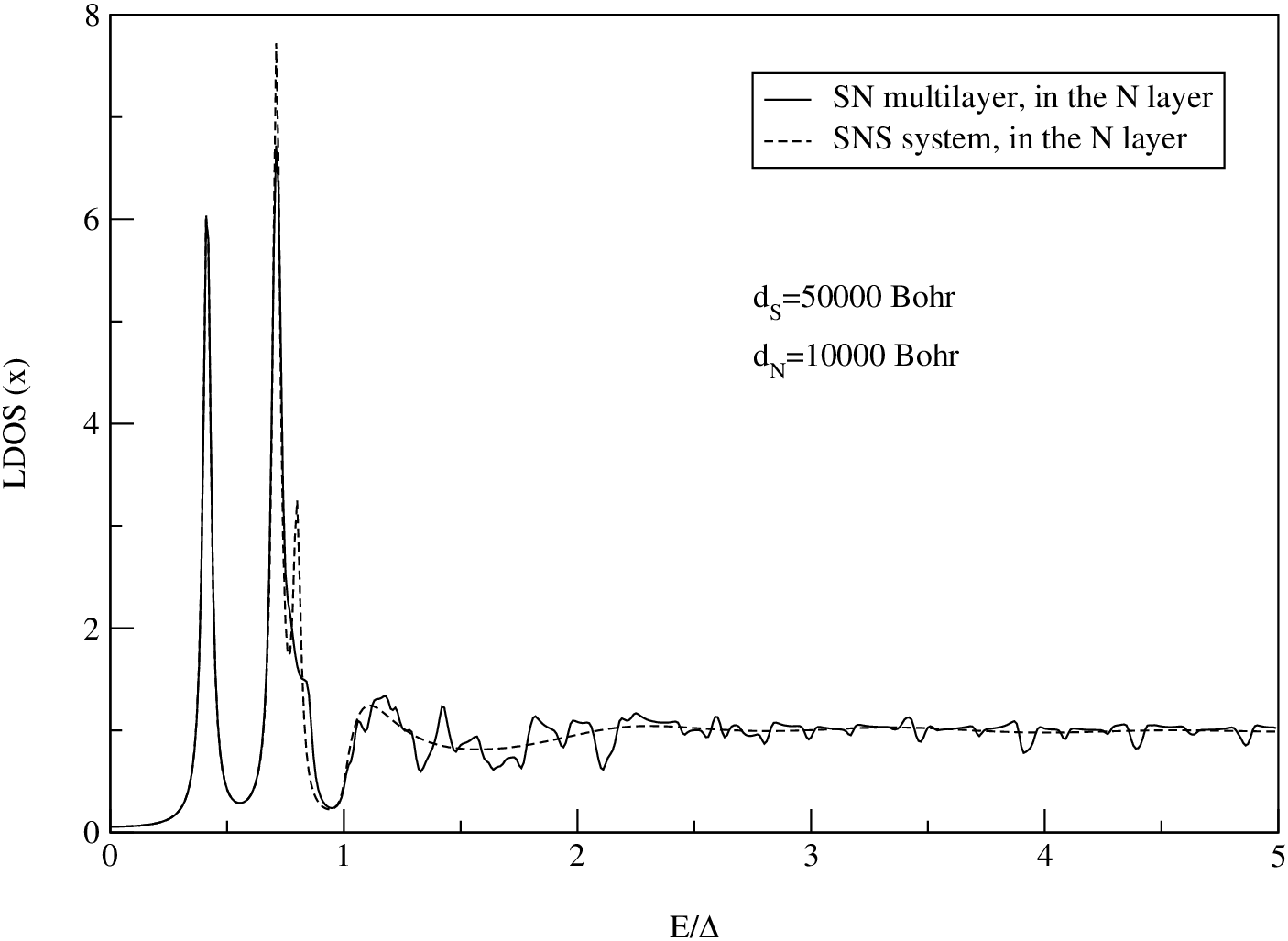,angle=0,width=7.1cm}}
\caption[]{LDOS for a SN multilayer ($d_S=50000$ Bohr and $d_N=10000$ Bohr) 
and for a SNS system (dashed line), calculated inside the N layer.}
\label{dN10dS50SNS}
\end{figure}

Let us pick up the N layer LDOS from Fig. \ref{dN10dS50} and put it 
together with the LDOS of a SNS system, whose $d_{N}=10000$ Bohr. 
This is what we show in Fig. \ref{dN10dS50SNS}. The good similarity 
is due to the large $d_S$. 
This SN multilayer is just a perturbation of a SNS system, as we can notice 
from the small oscillations at energies E which are larger than the gap. 
Way below $E=\Delta$ the Andreev bound states curves coincide. 
Just below $E=\Delta$ the clear peak in the SNS system curve is smeared 
out in the multilayer curve, due to tunneling interaction between the 
N-layers. 

\begin{figure}[htbp]
\centerline{\epsfig{figure=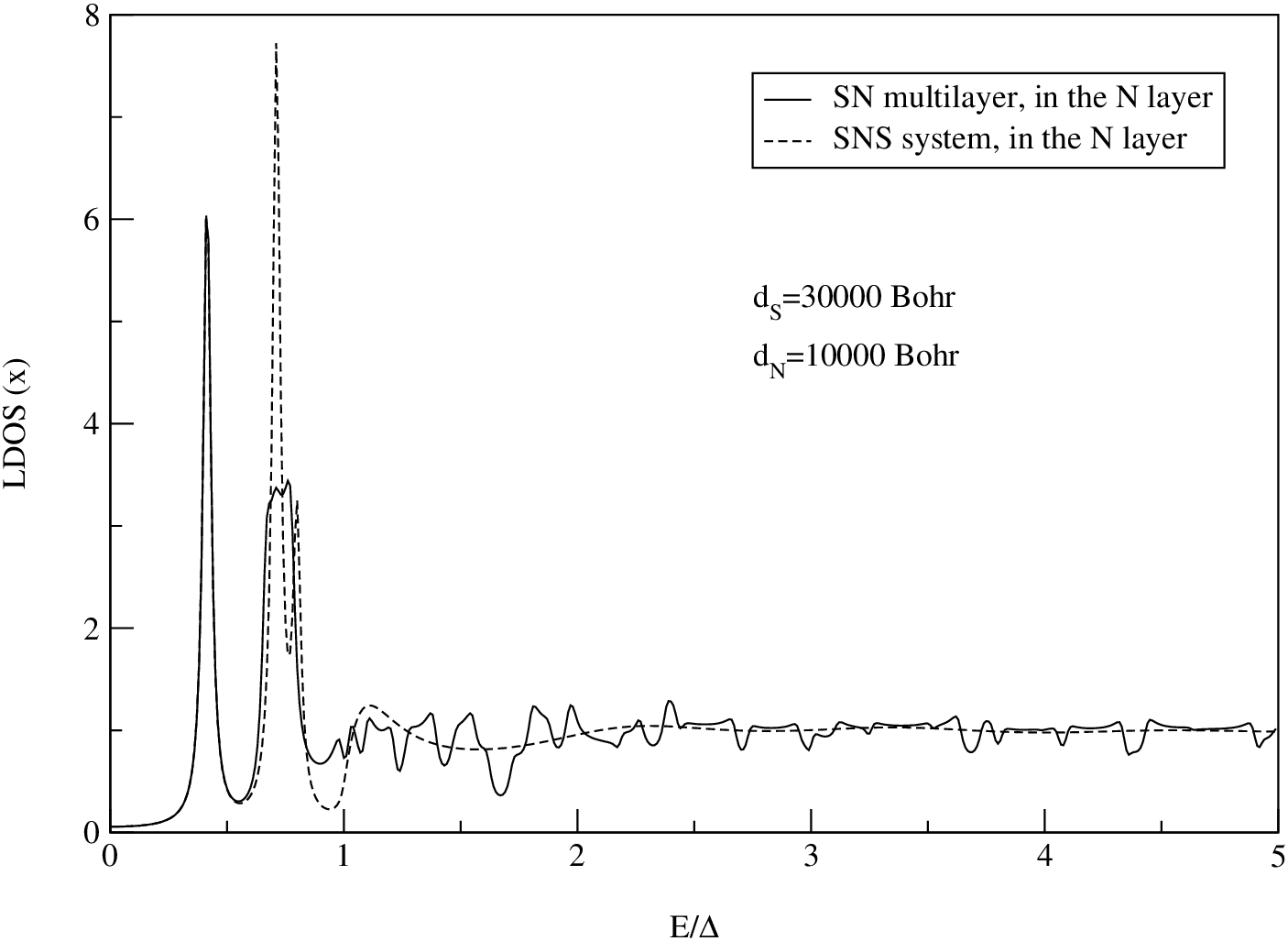,angle=0,width=7.1cm}}
\caption[]{LDOS for a SN multilayer ($d_S=30000$ Bohr and $d_N=10000$ Bohr) 
and for a SNS system (dashed line), calculated inside the N layer.}
\label{dN10dS30SNS}
\end{figure}

\begin{figure}[htbp]
\centerline{\epsfig{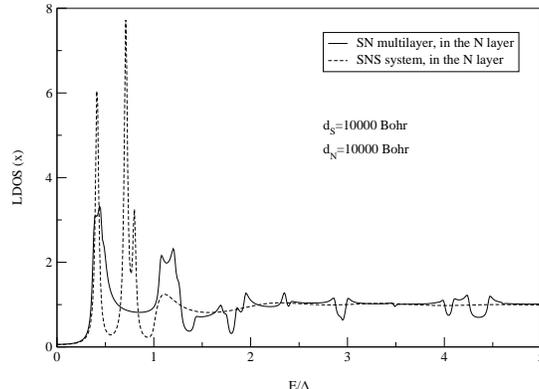}}
\caption[]{LDOS for a SN multilayer ($d_S=10000$ Bohr and $d_N=10000$ Bohr) 
and for a SNS system (dashed line), calculated inside the N layer.}
\label{dN10dS10SNS}
\end{figure}

The similarity between SNS systems and SN multilayers reduces 
with decreasing $d_{S}$. Multilayer 
features start to appear gradually in the LDOS, as we can see in Figs. 
\ref{dN10dS30SNS} and \ref{dN10dS10SNS}. 

Due to the increased tunnelling, the discrete 
states start to form bands, while at E larger than the gap, the oscillations 
are more pronounced and follow a periodicity, according to the dispersion 
relations \cite{6:Tanaka91} 
\begin{eqnarray}
\mbox{cos}[(k_x-k_{Fx})(d_S+d_N)]=
\mbox{cosh}\frac{\sqrt{E^2+|\Delta|^2}d_S}{2k_{Fx}}
\mbox{cosh}\Biggl(\frac{Ed_N}{2k_{Fx}}+\frac{i\phi}{2}\Biggr)
\nonumber \\
+\mbox{sinh}\frac{\sqrt{E^2+|\Delta|^2}d_S}{2k_{Fx}}
\mbox{sinh}\Biggl(\frac{Ed_N}{2k_{Fx}}+\frac{i\phi}{2}\Biggr)
\nonumber \\
\mbox{cos}[(k_x+k_{Fx})(d_S+d_N)]=
\mbox{cosh}\frac{\sqrt{E^2+|\Delta|^2}d_S}{2k_{Fx}}
\mbox{cosh}\Biggl(\frac{Ed_N}{2k_{Fx}}-\frac{i\phi}{2}\Biggr)
\nonumber \\
+\mbox{sinh}\frac{\sqrt{E^2+|\Delta|^2d_S}}{2k_{Fx}}
\mbox{sinh}\Biggl(\frac{Ed_N}{2k_{Fx}}-\frac{i\phi}{2}\Biggr), 
\label{dispersion}
\end{eqnarray}
where $k_{Fx}$ is given by 
\begin{equation}
k_{Fx}^2=\mu-\Biggl(\frac{n_y\pi}{L_t}\Biggr)^2
-\Biggl(\frac{n_z\pi}{L_t}\Biggr)^2.
\label{kFx}
\end{equation}

\begin{figure}[htbp]
\centerline{\epsfig{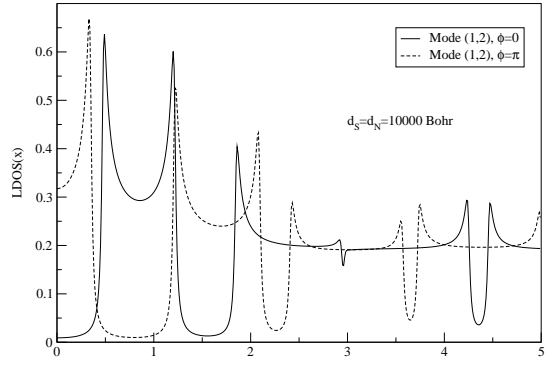}}
\caption[]{Contributions to the LDOS of a SN multilayer 
($d_S=d_N=10000$ Bohr) in the middle of the N layer 
from the mode (1,2), for two choices of the phase of the pair 
potential, $\phi=0$ and $\phi=\pi$.}
\label{mode12}
\end{figure}

In Fig. \ref{mode12} we illustrate these dispersion relations for 
a multilayer with $d_S=d_N=10000$ Bohr and two different choices for 
the phase difference, $\phi=0$ and $\phi=\pi$. 
To make it more clear, we considered here only the (1,2)-mode 
contribution to the LDOS. We notice the change in the succession 
of gaps with the phase $\phi$. 

\begin{figure}[htbp]
\centerline{\epsfig{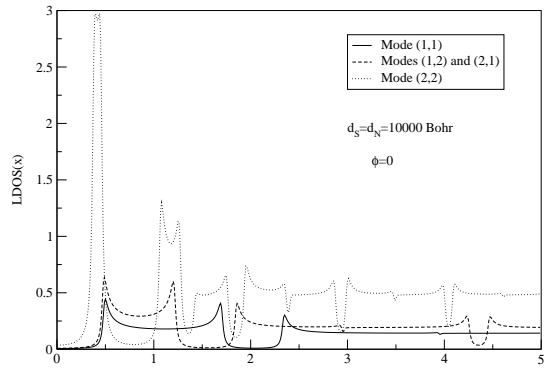}}
\caption[]{Contributions to the LDOS of a SN multilayer 
($d_S=d_N=10000$ Bohr) in the middle of the N layer 
from the modes (1,1), (1,2), (2,1), and (2,2). 
The phase of the pair potential is $\phi=0$.}
\label{modes}
\end{figure}

Using the definition of $k_{Fx}$, we notice that the number of allowed modes 
$(n_y,n_z)$ is limited by the condition $k_{Fx}^2\geq 0$. Besides, since 
$k_{Fx}$ is different for each mode,  the periodicity with which there 
is a solution for $k_x$ in the equations (\ref{dispersion}) is also different. 
For a higher mode $(n_y,n_z)$, $k_{Fx}$ is smaller and the frequency with 
which peaks and gaps in the LDOS are alternating increases. 
We can see this in Fig. \ref{modes}, where we compare the contributions 
to the LDOS 
coming from different modes, at phase $\phi=0$. As in Fig.  \ref{mode12}, 
the system has $d_S=d_N=10000$ Bohr. 
One can easily check out that by just adding the four modes' contributions 
one obtains the total LDOS shown in Fig. \ref{dN10dS10}, which we are 
going to discuss in the following subsection. 

\subsection{From infinite transverse size of a SN multilayer to a finite one}
\label{subsinf}

\begin{figure}[htbp]
\centerline{\epsfig{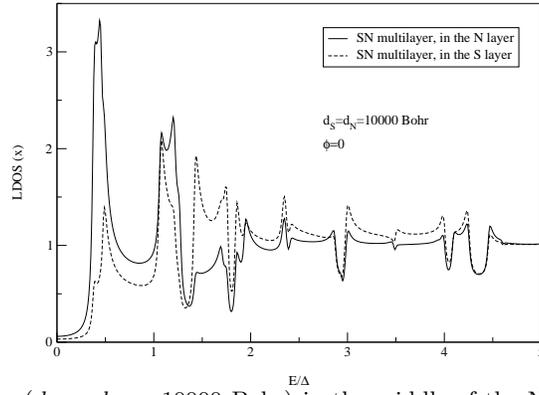}}
\caption[]{LDOS for a SN multilayer ($d_S=d_N=10000$ Bohr) 
in the middle of the N layer (solid line) and S layer (dashed line). 
The phase of the pair potential is $\phi=0$.}
\label{dN10dS10}
\end{figure}

\begin{figure}[htbp]
\centerline{\epsfig{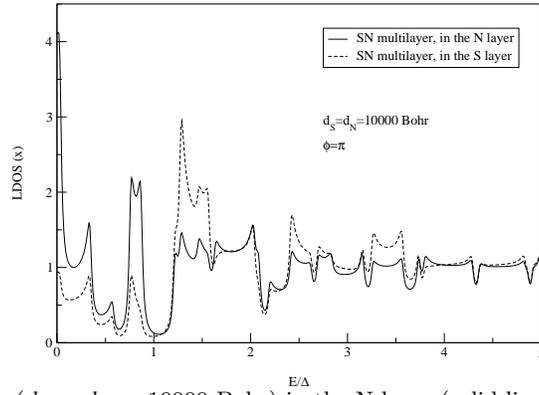}}
\caption[]{LDOS for a SN multilayer ($d_S=d_N=10000$ Bohr) 
in the N layer (solid line) and S layer (dashed line). The phase 
of the pair potential is $\phi=\pi$.}
\label{dN10dS10Pi}
\end{figure}

\begin{figure}[htbp]
\centerline{\epsfig{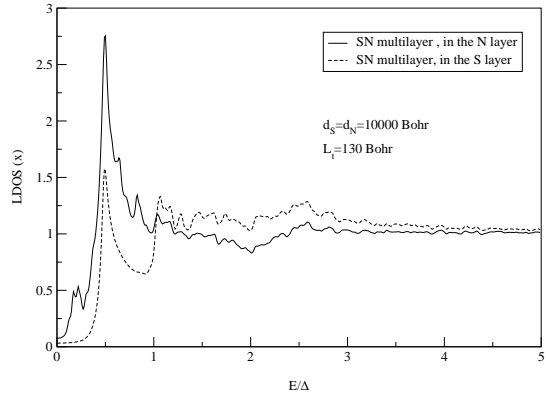}}
\caption[]
{LDOS calculated in the N and S layers of a SN multilayer, at a higher 
transverse size, $L_t=130$ Bohr. }
\label{SNLt}
\end{figure}

Now we are prepared to investigate real periodic-multilayer effects. 
In addition to the $\phi=0$ result shown in Fig. \ref{dN10dS10} we also show 
the $\phi=\pi$ result for the same system in Fig. \ref{dN10dS10Pi}. 
The dispersion relations (\ref{dispersion}) are nicely illustrated in the 
succession of bands and gaps. 
This system was 
studied also by Tanaka and Tsukada \cite{6:Tanaka91}. In their Fig. 2a, 
they describe a SN multilayer which has $d_N=d_S=$ 5000 \AA ~and $\phi=0$,  
and is infinite in the transverse direction. 
For $E<\Delta$ the pictures look quite similar, but above the gap the LDOS 
in Figs. \ref{dN10dS10} and \ref{dN10dS10Pi} is much less smooth than 
in Fig. 2 of Tanaka and Tsukada. 
However, in Fig. \ref{SNLt} we show the LDOS for a system with a larger 
transverse width ($L_t=130$ Bohr), and indeed, the behaviour for 
$E>\Delta$ has become much smoother. 
So we conclude that differences between our results and the results of 
Tanaka and Tsukada come from their use of an infinite transverse width. 

\section{Exact calculations at critical widths } 
\label{s:exact}

In many situations, the Andreev approximation gives results with an error 
which is estimated to be less than 0.1\%. The systems which we 
discussed up to now satisfy the conditions for which the Andreev 
approximation is very good. In this section, we will deal with situations 
in which exact calculations are necessary. 

\begin{figure}[htbp]
\centerline{\epsfig{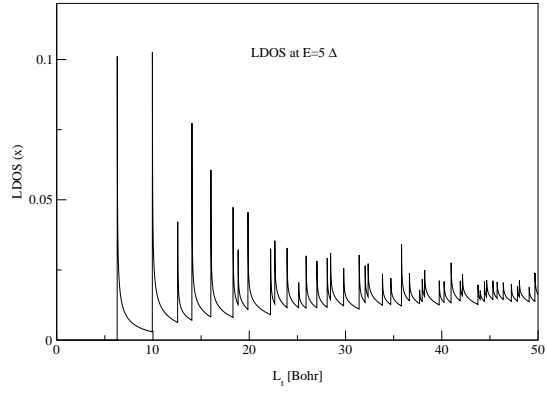}}
\caption[]
{LDOS calculated in a bar shaped S material, at $E=5\Delta$, 
at different transverse widths. }
\label{LDOSLt}
\end{figure}

In Fig. \ref{LDOSLt} we show the not-normalized LDOS at $E=5\Delta$ 
of a homogeneous bar as a function of the transverse 
width $L_t$. 
At $E=5\Delta$, the LDOS of multilayers with the same $L_t$ approaches 
a constant value, given by the homogeneous bar, as we can see in 
Figs. \ref{BulkS}~-~\ref{SNLt}. 
We notice in Fig. \ref{LDOSLt} that at certain values of $L_t$, 
the LDOS has steps, followed by a fast and smooth decrease. 
At these widths, where the condition $k_{Fx}^2\geq 0$ reaches the equality, 
an extra mode is allowed in addition to the previous ones. 
The new mode has a large contribution to the LDOS, explaining the step. 
Apparently, at smaller $L_t$'s the steps are higher, which means that 
the effect of adding a new mode is larger. 
These values of $L_t$ are called critical widths, and we will 
denote them by $L_t^{cr}$. 

Close to the critical widths, for $0\leq k_{Fx}^2\leq\Delta$, the 
AA is not good anymore. The highest modes contribute most to the 
LDOS, as the steps in Fig. \ref{LDOSLt} suggest us. Besides, since  
$k_{Fx}$ is very small in the dispersion relations (\ref{dispersion}), 
these modes will give rise to much more states than the lower modes. 

\begin{figure}[htbp]
\centerline{\epsfig{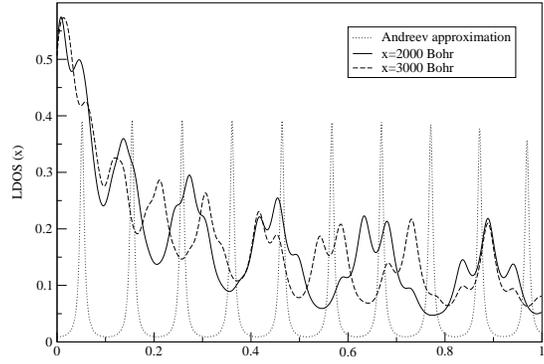}}
\caption[]
{LDOS calculated in a SN multilayer, for a width $L_t=12.566371$ Bohr close 
to a critical width $L_t^{cr}$, at different positions $x$ with respect to 
the S/N interface, inside the N layer. $d_N=d_S=10000$ Bohr. For comparison, 
Andreev approximation is represented with dotted line. } 
\label{LDOSx}
\end{figure}

As an illustration, in Fig. \ref{LDOSx} we show the absolute value 
of the LDOS for a SN multilayer, whose $d_S=d_N=10000$ Bohr, 
(like we showed in Fig. \ref{dN10dS10}), 
but this time at a width $L_t=12.566371$ Bohr. 
This value of the transverse width, corresponding to the equality 
$k_{Fx}^2=0.3\Delta$ for the highest, (2,2) mode, 
lies very close to the critical width corresponding to $k_{Fx}^2=0$ 
and can be implemented without getting numerical problems. 
At this width, we calculated the LDOS at two different positions $x$ with 
respect to the SN interface, inside the N layer. Although the solid and 
dashed curves have peaks at the same energies (the dispersion relations 
do not change with $x$), their magnitude goes up or down, depending very 
much on $x$. 
This is not the case within the Andreev approximation, represented 
for comparison with a dotted line, where peaks of the same height are 
situated on the energy axis at equal distance from each other. 
This difference between the exact and the approximate results 
can be explained if we make use of the definition of the LDOS, 
\begin{equation}
LDOS({\mathbf r})=\sum_n|\Psi_n({\mathbf r})|^2\delta(E-E_n). 
\label{ldos}
\end{equation}
In the AA the Andreev states with the electron moving to the right and 
to the left are uncoupled and are degenerate. 
They can be represented by plane waves, having a ${\mathbf r}$-independent 
absolute value. 
In the exact treatment the corresponding travelling waves are coupled and 
they are split into two standing waves, 
an odd (sinus) and even (cosinus) function. 
This leads to weighting factors in the expression of the LDOS which are 
different and position-dependent. 
So, the lifted degeneracy in the exact calculation explains the position 
dependence of the LDOS illustrated in Fig. \ref{LDOSx}. 

\begin{figure}[htbp]
\centerline{\epsfig{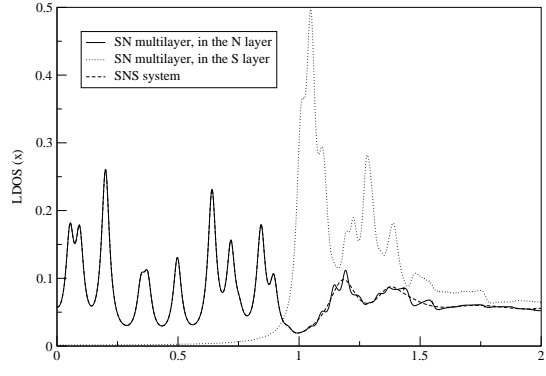}}
\caption[]
{LDOS calculated in both S and N layers of a SN multilayer, with 
$d_N=d_S=4000$ Bohr, as well as 
in a SNS system, at a transverse width $L_t=12.5676$ Bohr. }
\label{SNmSNS}
\end{figure}

\begin{figure}[htbp]
\centerline{\epsfig{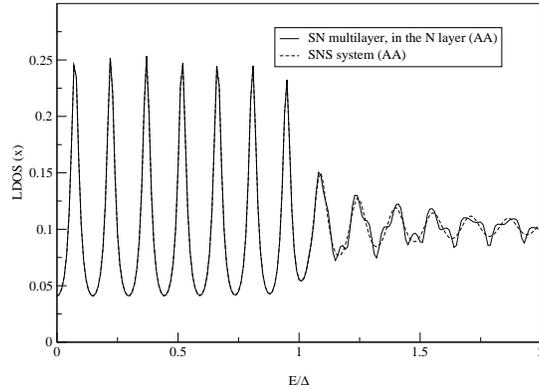}}
\caption[]
{LDOS calculated in Andreev approximation in the N layer of a SN multilayer, 
with $d_N=d_S=4000$ Bohr, as well as 
in a SNS system, at a transverse width $L_t=12.5676$ Bohr. }
\label{LDOSAAcrit}
\end{figure}

In order to compare multilayer results with publicated SNS results 
\cite{roland-th,roland2}, we calculated the LDOS inside the N and S layers 
of a multilayer with $d_N=d_S=4000$ Bohr. This is shown in 
Fig. \ref{SNmSNS} together with the comparing SNS result. 
The corresponding results derived in the Andreev approximation 
are shown in Fig. \ref{LDOSAAcrit}. 
We restricted the calculations to the highest mode's 
(2,2) contribution to the LDOS, at $L_t=12.5676$ Bohr. 
This transverse width corresponds to $k_{Fx}^2=\Delta$ for the mode (2,2). 
In both Figures, at energies $E<\Delta$, we do not notice 
any difference for the N-layer. 
The multilayer features appear only above the gap. Besides, 
inside the S-layer there is no contribution from the highest mode, as the 
corresponding states have such a small momentum $k_x$ 
that, for $E<\Delta$, they are localized inside the N layer. 

The features shown in this section are directly related to a fine-tuning 
of the transverse width. 
In this respect, these results are new compared to those reported by Tanaka 
and Tsukada \cite{6:Tanaka91}, who have considered an infinite transverse 
width only. 

\section{Calculation of the supercurrent} 
\label{s:currents} 

This section is devoted to the supercurrent in a SN multilayer. 
In 1962 Josephson predicted that a supercurrent can be present in a 
SIS junction (Josephson junction) in the absence of an external voltage 
(dc Josephson effect). 
This current appears provided there is a difference $\phi$ in the phase of the 
pair potential between the two S layers of the Josephson junction. 
\begin{equation}
I=I_{max}\mbox{sin}\phi. 
\end{equation} 
Further, if an external potential is applied to the junction, then 
\begin{equation}
\frac{d\phi}{dt}=2eV/\hbar. 
\end{equation}
In other words, an external potential gives rise to an alternating 
supercurrent of frequency $f=2eV/h$ (so called ac Josephson effect). 
The quantum energy $hf$ equals the energy of a Cooper pair transferred across 
the junction. 
It appears that the Josephson effect is present also in SNS junctions. 
We will investigate it for SN multilayers. 

\begin{figure}[htbp]
\centerline{\epsfig{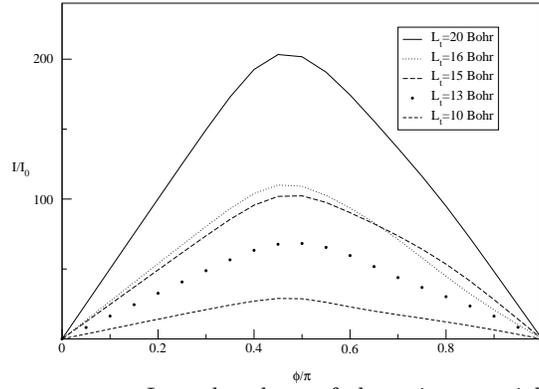}}
\caption[]
{The dependence of the supercurrent $I$ on the phase of the pair 
potential $\phi$. $d_N=d_S=10000$ Bohr and the normalization factor 
$I_0=e\Delta /\hbar$. }
\label{StroomPhase}
\end{figure}

Using equation (\ref{cs:current}), we calculated the supercurrent $I$ through 
a SN multilayer with $d_S=d_N=10000$ Bohr as a function of 
the phase difference $\phi$ between two consecutive S layers. 
Fig. \ref{StroomPhase} gives the supercurrent normalized to the 
basic supercurrent unit $I_0=e\Delta/\hbar$ for  different choices 
of the transverse width $L_t$ of the multilayer. 
The $\phi$ dependence of $I$ is basically similar to a sin$\phi$ 
dependence, as for a Josephson junction, in that it is periodic 
in $2\pi$. 
We notice that the supercurrent increases in magnitude with the 
transverse width, which makes sense, given the fact that the larger 
the width $L_t$, the more modes contribute to the current. 
However, a small deviation from this monotonic behaviour in the dependence 
of the supercurrent on the transverse widths is noticed 
at a larger phase, $\phi\approx2\pi/3$. In our Fig. \ref{StroomPhase} 
we see this behaviour between the critical widths $L_t^{cr}=14.0492$, when 
the mode (3,1) starts to contribute and $L_t^{cr}=16.0186$, 
when the mode (3,2) appears. 
Around $\phi=2\pi/3$, the curve corresponding to 
$L_t=$15 Bohr lies slightly higher than the curve for $L_t=$16 Bohr. 
This can be interpreted as being due to a destructive interference between the 
electronic contributions to the current at larger phase. 
This results in a small deviation from the symmetry of the sin-function 
dependence of the supercurrent as a function of phase. 
For an infinite transverse width, Tanaka and Tsukada show a similar 
dependence in Fig. 4 of their paper \cite{6:Tanaka91}. 

\begin{figure}[htbp]
\centerline{\epsfig{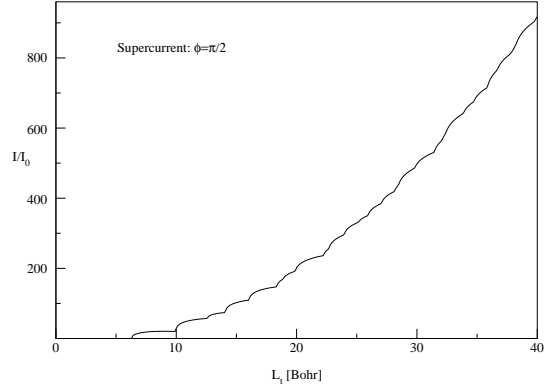}}
\caption[]
{The dependence of the supercurrent $I$ on the transverse length $L_t$. 
$I_0=e\Delta /\hbar$. }
\label{StroomLt}
\end{figure}

The way in which the transverse width influences the maximum of the 
supercurrent $I_{max}$ is shown in Fig. \ref{StroomLt}. 
The monotonic increase of the supercurrent 
exhibits steps at each critical width. This is not surprising, 
since at a critical width new modes start to contribute. 
At the onset of this contribution, the kinetic energy of the new modes 
$k_{F}^2-(\frac{n_{y,max}\pi}{L_t})^2-(\frac{n_{z,max}\pi}{L_t})^2$ is 
very small and so is their contribution to the supercurrent. 
But with the increase of $L_t$, the supercurrent reaches a constant 
regime, till the next $L^{cr}_{t}$. 

\begin{figure}[htbp]
\centerline{\epsfig{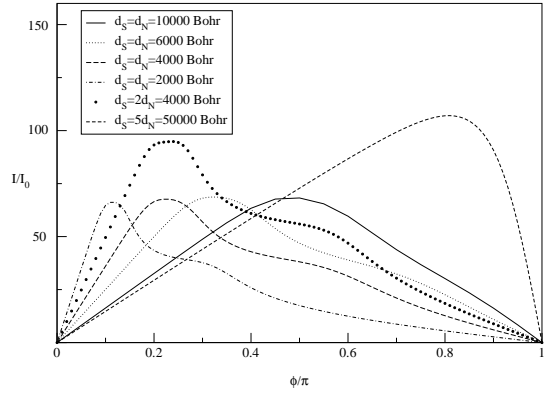}}
\caption[]
{The dependence of the supercurrent $I$ on $d_N$ and $d_S$ for $L_t$=13 Bohr. 
$I_0=e\Delta /\hbar$. }
\label{StroomdNdS}
\end{figure}

Now we fix the transverse width at $L_t=13$ Bohr and we change the 
layer thicknesses $d_S$ and $d_N$. The results are shown in Fig. 
\ref{StroomdNdS}. 
If $d_S\gg\xi$, in which the coherence length $\xi\approx 4000$ Bohr, 
the SN multilayer compares well to a SNS system, for which 
$\phi_{max}\approx 0.8\pi$, as we will see below in discussing 
Fig. \ref{StroomSNS}. 
At smaller values of $d_S$ 
the phase $\phi_{max}$ at which the current has a maximum shifts 
gradually towards lower values. 
Further, if the ratio between $d_S$ and $d_N$ is constant, the systems have 
approximatively the same maximum supercurrent. 
However, it should be noticed that all systems have the same $\Delta$. 
This picture of constant $I_{max}$ changes if the gap function is calculated 
selfconsistently, as will be shown in Section \ref{s:selfcons}. 
With the decrease of $d_N$ with respect to $d_S$, 
the current increases due to a better coupling between 
the S layers. 
Modifying the value of $d_N$ doesn't have consequences on $\phi_{max}$. 
This can be noticed if we compare the curves corresponding 
to $d_N=d_S=4000$ Bohr and $d_N=d_S/2=2000$ Bohr.

\section{Selfconsistent calculation of the gap }
\label{s:selfcons}

The formalism described in Section \ref{s:theory} can be applied 
to a selfconsistent calculation of the gap function $\Delta$. The method, 
which is extensively described in Refs. \cite{roland-th,roland2}, 
is based on the selfconsistency condition 
\begin{equation}
\Delta(x)=-k_BTV(x)\frac{1}{L_t^2}\sum_{k_y,k_z>0}\sum_{n}
G_{12}(x,x;k_y,k_z,i\omega_n). 
\label{selfc}
\end{equation}

\begin{figure}[htbp]
\centerline{\epsfig{figure=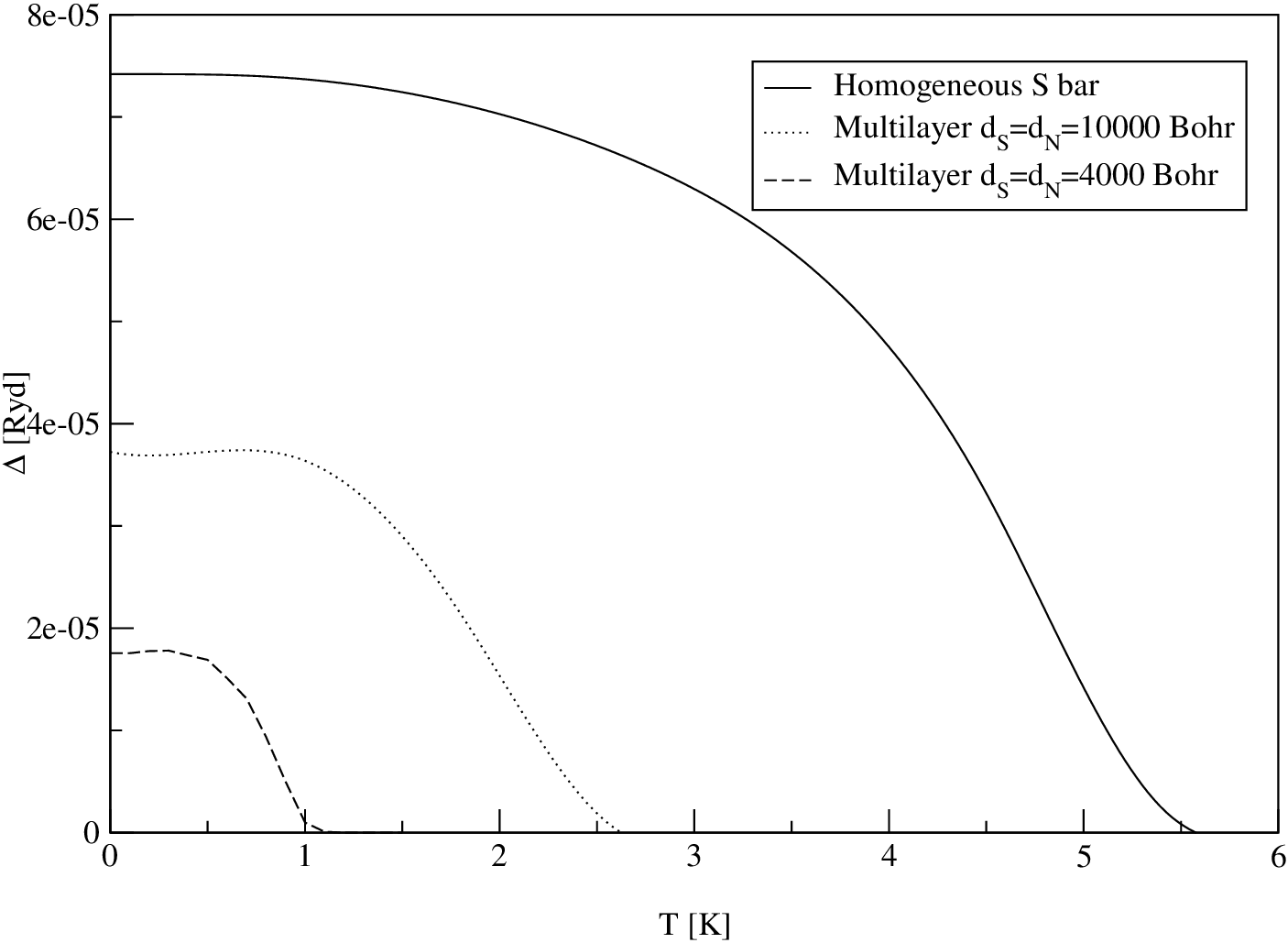,angle=-90,width=7.1cm}}
\caption[]
{Selfconsistent gap calculations for systems with transverse width 
$L_t$=30 Bohr. }
\label{DelT30}
\end{figure}

\begin{figure}[htbp]
\centerline{\epsfig{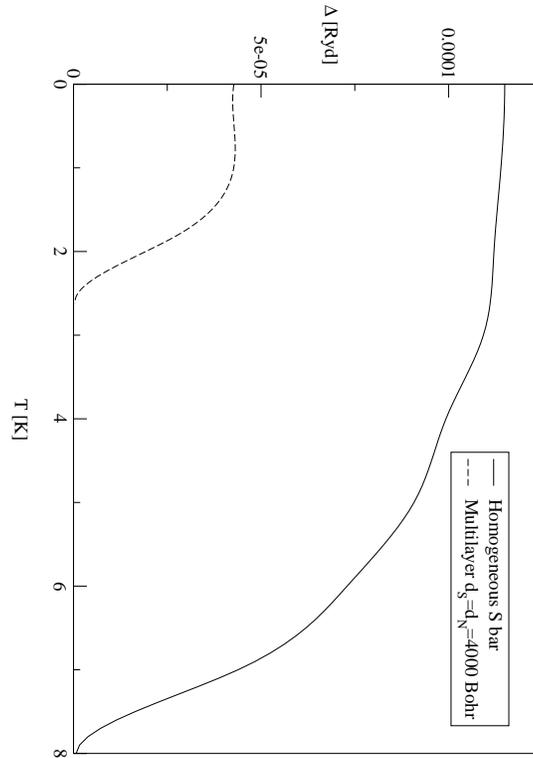}}
\caption[]
{Selfconsistent gap calculations for systems with transverse width 
$L_t$=100 Bohr. }
\label{DelT100}
\end{figure}

The summation in equation (\ref{selfc}) is divergent. 
In order to render the sum convergent, a cut-off of the summation over 
the Matsubara frequencies is introduced, as in the following expression 
\begin{equation}
\Delta(x)=-k_BTV(x)\frac{1}{L_t^2}\sum_{k_y,k_z>0}
\sum_{|\omega_n|=\pi k_BT}^{\omega_D}
G_{12}(x,x;k_y,k_z,i\omega_n), 
\label{cutoffn}
\end{equation}
where $\omega_D$ is the Debye frequency. 
We limit the Matsubara frequency 
to $\omega_D=n_{max}\pi k_BT$, where $n_{max}=[\Theta_D/\pi T]$ 
and $\Theta_D=\omega_D/k_B$ the Debye temperature. 
As we notice, at large temperatures, $n_{max}$ 
becomes small, while $d\omega_n=\omega_n-\omega_{n-1}$ is large. 
This gives rise to big, unphysical oscillations of the order 
parameter $\Delta$ with temperature $T$, close to $T_c$. 
We get rid of these unwanted oscillations by taking an integration, 
rather than a summation over $\omega_n$. 
Results for a bar of transverse widths $L_t=$ 30 and 100 are shown in 
Fig. \ref{DelT30} and Fig. \ref{DelT100} respectively, at which we will 
come back later in this section. 

\begin{figure}[htbp]
\centerline{\epsfig{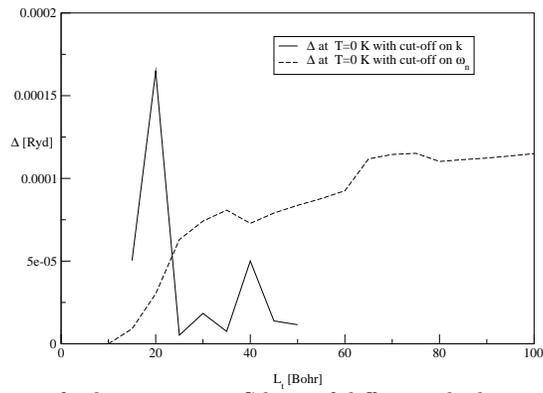}}
\caption[]
{Selfconsistent gap calculations for homogeneous S bars of different 
thicknesses $L_t$, at $T=0$, for two different cut-off methods. } 
\label{DeltaLtT0}
\end{figure}

In addition to the integration over the Matsubara frequencies, 
we also investigated 
another cut-off method, which avoids the gap oscillations at large 
temperatures. 
An alternative way to render the summation (\ref{selfc}) convergent, 
is to impose the cut-off on the momenta $k$, instead of on the 
Matsubara frequencies. 
For a homogeneous S bar this reads 
\begin{equation}
\Delta(x)=-\frac{k_BTV(x)}{8\pi}\frac{1}{L_t^2}
\int_{k^2=\mu-k_B\Theta_D}^{\mu+k_B\Theta_D}{dk_x\sum_{k_y,k_z} \sum_{n}
G_{12}(k_x,k_y,k_z,i\omega_n)}. 
\label{cutoffk}
\end{equation}
More explicitly, equation (\ref{cutoffk}) can be written 
\begin{equation}
\Delta(x)=-\frac{k_BTV(x)}{8\pi}\frac{1}{L_t^2}
\int_{k^2=\mu-k_B\Theta_D}^{\mu+k_B\Theta_D}{dk_x\sum_{k_y,k_z} \sum_{n}
\frac{\Delta}{(i\omega_n)^2-(\mu-k_x^2-k_y^2-k_z^2)^2-\Delta^2}}.
\label{cutoffk1}
\end{equation}
We further perform the summation over the Matsubara frequencies, obtaining 
\begin{equation}
\Delta(x)=-\frac{V(x)\Delta}{8\pi}\frac{1}{L_t^2}
\int_{k^2=\mu-k_B\Theta_D}^{\mu+k_B\Theta_D}{dk_x\sum_{k_y,k_z} 
\frac{\mbox{tanh}\frac{E_k}{2k_BT}}{2E_k}}, 
\label{cutoffk2}
\end{equation}
where $E_k^2=(\mu-k_x^2-k_y^2-k_z^2)^2+\Delta^2$. 
For bulk superconductors, both ways of rendering the integral convergent 
lead to the same result. 
However, in the case of a homogeneous superconducting bar, 
we have summation 
over the transverse momenta $k_y$ and $k_z$ instead of an integration. 
This affects dramatically the results when $k_y$ and $k_z$ are large, 
particularly at small $L_t$. We can see this in the dependence 
of the gap $\Delta$ on the transverse width $L_t$, shown in 
Fig. \ref{DeltaLtT0}. The calculation is done at $T=0$. 
The dotted line comes from a calculation with a cut-off on the Matsubara 
frequencies. Apparently, the latter cut-off method leads to a much more 
stable result than the method in which the momenta are cut off. 
The solid line exhibits unphysical oscillations in the gap indeed. 
In addition we show a dependence $\Delta (T)$ in 
Fig. \ref{DeltaTLt30}. 
This curve exhibits a largely reduced superconductivity compared to the upper 
curve in Fig. \ref{DelT30} obtained by the other cut-off method. 
We conclude that cutting off the Matsubara frequencies leads to much more 
reliable results. 
Otadoy {\it et al.} \cite{roland-th,roland2} used this method to calculate 
the selfconsistent gap for systems such as SNS and SNSNS systems. 
Here, we extend the application to SN multilayers. We show results for two 
transverse widths. 

In Fig. \ref{DelT30} the temperature dependence of 
the gap is shown for $L_t$=30 Bohr. 
The solid, dotted, and dashed curves represent $\Delta(T)$ for a homogeneous 
S bar, a SN multilayer with $d_S=d_N=10000$ Bohr, and a SN multilayer with 
$d_S=d_N=4000$ Bohr respectively. 
As expected, the selfconsistent gap decreases with the periodicity 
$d_S+d_N$ of the multilayer. 
Indeed, for a smaller $d_S$, the contribution to the averaged gap 
over the layer comes mostly from the regions close to the NS interface, 
where the suppression of the gap is most effective. 
Similarly, in Fig. \ref{DelT100} we show results for systems with 
$L_t$=100 Bohr. 
Again, one clearly sees the suppression of superconductivity by reducing the 
transverse width. 

In addition to the results derived in Section \ref{s:currents}, in the 
present stage we can look to the temperature dependence of the supercurrent. 
First we show in Fig. \ref{StroomNb} the phase dependence of the 
supercurrent for multilayers with $L_t=30$, at different temperatures. 
For a given layer thickness, 
the peaks at different temperature occur at the same phase. 
The multilayer with a smaller periodicity $d_S+d_N$ has the corresponding 
maximum at a lower $\phi$, as we discussed in the previous section. 
We notice the suppression of the current with the increase of $T$. 
For the multilayer with $d_S=d_N$=10000 and $L_t$=30, we show the 
current-temperature dependence $I(T)$, in Fig. \ref{it}. 
The temperature at which the superrcurent becomes zero coincides with 
the critical temperature at which the corresponding gap function is zero, 
see Fig. \ref{DelT30}. 

Finally, we compare the phase dependence of the supercurrent 
for SN multilayers with corresponding results for the SNS system. 
In Fig. \ref{StroomSNS} we show results at temperature $T=0.4$ K. 
Obviously, since the gap function in the SNS systems is equal to the 
homogeneous bar gap, the supercurrent is larger than in the SN multilayers. 
However, with the increase of the N-layer thickness, the supercurrent 
in the SNS system decreases, since to a thicker N-layer corresponds 
a weaker coupling between the two half-infinite S-layers. 
On the contrary, for the SN multilayer results shown, the S-layer thickness 
increases as well when the N-layer thickness is increased, and the 
supercurrent increases. 
In consistency with the discussion of Fig. \ref{StroomdNdS}, the phase at 
which the supercurrent has a maximum, $\phi_{max}$ does not depend on 
the N-layer thickness, and it shifts to the right with increasing $d_S$. 

\section{Interface potentials}
\label{s:interface}

Stimulated by recent work on the influence of interface barriers in SNS 
systems \cite{zaikin}, we studied this in more detail and for SN multilayers 
as well. 
Interface barriers can come out in practice as an effect of localized 
disorder at the interface or as a typical oxide layer in a point contact. 

A simple model of a $\delta$-function potential at the interfaces introduced 
by Blonder {\it et al.} \cite{blonder}, can be implemented in our formalism 
easily. 
The corresponding Hamiltonian for interfaces at positions $x_j$ reads as 
\begin{equation}
{\cal H}_x\equiv-\frac{d^2}{dx^2}-k_{Fx}^2+\sum_jW_j\delta(x-x_j), 
\end{equation}
where $W_j$ is the strength of the barrier and can be estimated 
using the transmission coefficient of the barrier 
\begin{equation}
{\cal T}\equiv\frac{1}{1+mW^2/\hbar^2 E}. 
\end{equation} 
In the presence of a $\delta$-function barrier, the wave function is still 
given by equation (\ref{wavefct}), but the boundary conditions for the 
Green's function read now 
\begin{equation}
\sum_{\nu}S_{\nu j}G_{\nu j\nu'j}(x_j,x')=0, 
\end{equation}
where 
\begin{equation}
\label{snj}
S_{\nu j}=\left( \begin{array}{cccc}
 ~~\nu & ~~0 & ~~0 & ~~0 \\
 ~~0 & ~~\nu & ~~0 & ~~0 \\
 ~~-\frac{1}{2}W_j & ~~0 & ~~\nu & ~~0 \\
 ~~0 & ~~-\frac{1}{2}W_j & ~~0 & ~~\nu 
\end{array} \right). 
\end{equation}
We choose $W_j=W$ the same for each SN interface. 

Before applying this to SN structures, 
we first look at the bound states of a SNS system at different 
choices of the barrier strength W. In Fig. \ref{potintSNS} 
we show the LDOS of a SNS system characterized by $d_N=10000$ 
Bohr and $L_t=13$ Bohr. 
In the absence of the interface potential, the SNS system LDOS has four 
bound states for $E<\Delta$, corresponding to the modes (1,1), (1,2), (2,1), 
and (2,2). 
Due to the fact that $L_y=L_z=L_t$, the states corresponding to the 
(1,2) and (2,1) modes are degenerate. 
Apparently, as we can notice, the presence of the $\delta$-function potential 
favours the appearance of new bound states, due to the scattering with 
the interface potential. This implies that in the presence of a 
scattering potential even in AA the bound states split up. 

In the case of the multilayer, for which results are shown in 
Fig. \ref{potintLDOS}, the deviation from the zero potential case 
is even more pronounced, as the quasi-periodicity of the dispersion 
relations (\ref{dispersion}) 
is perturbed by the interface barrier. As for the SNS system, 
new bound states appear and complicate the picture seen in 
Fig. \ref{dN10dS10}, which describes the same multilayer, but in 
the absence of an interface barrier. 

In the limit of a large barrier strength $W$ ($W>1$ RydBohr), the S-layers 
decouple, so that the density of states for a multilayer becomes similar 
to the one of a SNS system. 
This can be seen in Fig. \ref{W10}, in which we show the LDOS for a SN 
multilayer with $d_S=d_N=10000$ and $L_t=13$, and for a SNS system with 
$d_N=10000$. 
The barrier strength is $W=10$ RydBohr. 
Compared to the strength of a S layer, which is $\Delta d_S=1$ RydBohr, 
this interface barrier is 10 times larger. 
At such a large strength of the barrier, in the N-layer there are just 
bound states. 
To make it more clear, in Fig. \ref{W10modes} we show the contribution from 
each transverse mode to the LDOS of a SNS system, for energies up to 
10 times the gap. 
The peaks corresponding to the same mode $(n_y,n_z)$ have the same height 
and obey the dispersion relation for a 3-dimensional box, 
\begin{equation}
E+\mu=k_x^2+k_y^2+k_z^2
=\Biggl(\frac{(2n_x+1)\pi}{d_N}\Biggr)^2
+\Biggl(\frac{n_y\pi}{L_t}\Biggr)^2
+\Biggl(\frac{n_z\pi}{L_t}\Biggr)^2, 
\label{3dbox}
\end{equation}
with $2n_x+1\ge \frac{k_{Fx}d_N}{\pi}$ and 
$k_{Fx}=\sqrt{\mu-(\frac{n_y\pi}{L_t})^2-(\frac{n_z\pi}{L_t})^2}$. 
Thus, for the (1,1) mode, the first peak has $n_x=985$ and is situated 
at the energy $E=2\Delta$, while the second peak has $n_x=986$ and occurs 
at $E=9.7\Delta$. 
Similarly, using Eq. (\ref{3dbox}) for the modes (1,2) and (2,1), we obtain 
peaks for $n_x=726$ at $E=3.6\Delta$ and for $n_x=727$ at $E=9.3\Delta$. 
For the mode (2,2) we get peaks for $n_x=288, 289, 290, 291, {\rm and~} 292$, 
at the energies $E=0.6\Delta, 2.9\Delta, 5.1\Delta, 7.4\Delta, {\rm and~}
9.7\Delta$ respectively. 

The fact that the S-layers decouple in the limit of large barrier strength 
has also consequences on the phase dependence of the LDOS. 
We first show in Fig. \ref{W0ph} the LDOS of an SN multilayer without 
interface barrier for $\phi=0$ and $\phi=\pi$. 
In this figure the solid curves of the Figs. \ref{dN10dS10} and 
\ref{dN10dS10Pi} are shown in one picture. 
Clearly, the features of the LDOS, already discussed in Section 
\ref{s:andreev}, are different for the two values of the phase $\phi$. 
However, in the presence of an interface barrier, the picture changes. 
In Figs. \ref{W1ph} and \ref{W10ph} we show the LDOS for $W=1$ and $W=10$ 
respectively. 
When $W=1$, the bound states occur almost at the same energies for both 
phases, and for $W=10$ the LDOS almost coincide, as a result of a complete 
decoupling of the successive S-layers. 
This leads to a total suppression of the supercurrent at large values 
of $W$. 

The calculated supercurrent $I$ for a SN multilayer at different barrier 
strengths $W$ is shown in Fig. \ref{potintI}. 
Clearly, an interface barrier diminishes the supercurrent. 
The stronger the barrier, the smaller is the transmission probability through 
the interface. 
For values of $W$ larger that 2 RydBohr the supercurrent is completely 
suppressed. 
A similar result was obtained in the recent study mentioned above, of the 
Josephson current in the much simpler SNS system having several insulating 
barriers \cite{zaikin}. 

\section{Conclusions}
\label{s:conclusion}

In this paper we discussed SN multilayer structures. In particular 
we show results for periodic infinite multilayers, represented by 
a Kronig-Penney superlattice model. 
By applying a Green's function 
formalism, we focussed first on the Andreev bound states 
and we studied the limitations of the 
Andreev approximation in relation to the finite transverse size of the 
systems. 

Further, we calculated the supercurrent through such a periodic SN multilayer. 
We completed this study using a selfconsistently calculated gap. 
Finally, including a $\delta$-function potential at the interface, we 
derived results which account for possible barrier scattering at the 
interfaces. 

The results presented in this paper are meant to understand 
the physics which is behind SN multilayer structures. 
For our purpose it is more appropriate to investigate systems of very 
small transverse size because in such systems the effects of the breakdown 
of the Andreev approximation come out most clearly. 
However, at the moment 
there are no experimental data to which we can compare. 
For larger systems, more accessible to experiments, 
the physics remains the same, but their complexity could obscure some 
of the fundamental aspects we are looking at. 

Applications to intrinsic Josephson junctions \cite{muller} made from 
high-T$_c$ superconducting materials would require an extension of the 
present theory to the case of d-wave symmetry of the order parameter. 

\begin{appendix}
\section*{The matrices $\hat{\mathbf A}$, $\hat{\mathbf B}$, 
and $\hat{\mathbf C}$} 
\label{appendix}

In this Appendix we show the structure of the equations (\ref{def:Xminus}) 
to (\ref{def:Tjj}) more explicitly. 
The matrices $\hat{\mathbf A}$ and $\hat{\mathbf C}$ are highly singular 
and due 
to this, the  solutions of Eqs. (\ref{def:Xminus}) and (\ref{def:Xplus}) 
are sparse matrices, which can be written 
\begin{equation}
\hat{\mathbf X}_{-}=
\left( \begin{array}{cccc}
~~\hat{0} & ~~\hat{0} & ~~\hat{x}_-^{13} & ~~\hat{0} \\
~~\hat{0} & ~~\hat{0} & ~~\hat{x}_-^{23} & ~~\hat{0} \\
~~\hat{0} & ~~\hat{0} & ~~\hat{x}_-^{33} & ~~\hat{0} \\
~~\hat{0} & ~~\hat{0} & ~~\hat{x}_-^{43} & ~~\hat{0} \\
\end{array} \right)=
\left( \begin{array}{cccc}
~~\hat{0} & ~~\hat{0} & 
~~\left( \begin{array}{cc}
X_-^{15} & X_-^{16} \\
X_-^{25} & X_-^{26} \\
\end{array} \right) 
& ~~\hat{0} \\
~~\hat{0} & ~~\hat{0} & 
~~\left( \begin{array}{cc}
X_-^{35} & X_-^{36} \\
X_-^{45} & X_-^{46} \\
\end{array} \right) 
& ~~\hat{0} \\
~~\hat{0} & ~~\hat{0} & 
~~\left( \begin{array}{cc}
X_-^{55} & X_-^{56} \\
X_-^{65} & X_-^{66} \\
\end{array} \right) 
& ~~\hat{0} \\
~~\hat{0} & ~~\hat{0} & 
~~\left( \begin{array}{cc}
X_-^{75} & X_-^{76} \\
X_-^{85} & X_-^{86} \\
\end{array} \right) 
& ~~\hat{0} \\
\end{array} \right) 
\end{equation}
and 
\begin{equation}
\hat{\mathbf X}_{+}=
\left( \begin{array}{cccc}
~~\hat{0} & ~~\hat{0} & ~~\hat{0} & ~~\hat{x}_+^{14} \\
~~\hat{0} & ~~\hat{0} & ~~\hat{0} & ~~\hat{x}_+^{24} \\
~~\hat{0} & ~~\hat{0} & ~~\hat{0} & ~~\hat{x}_+^{34} \\
~~\hat{0} & ~~\hat{0} & ~~\hat{0} & ~~\hat{x}_+^{44} \\
\end{array} \right)=
\left( \begin{array}{cccc}
~~\hat{0} & ~~\hat{0} & ~~\hat{0} & 
~~\left( \begin{array}{cc}
X_+^{17} & X_+^{18} \\
X_+^{27} & X_+^{28} \\
\end{array} \right) \\
~~\hat{0} & ~~\hat{0} & ~~\hat{0} & 
~~\left( \begin{array}{cc}
X_+^{37} & X_+^{38} \\
X_+^{47} & X_+^{48} \\
\end{array} \right) \\
~~\hat{0} & ~~\hat{0} & ~~\hat{0} & 
~~\left( \begin{array}{cc}
X_+^{57} & X_+^{58} \\
X_+^{67} & X_+^{68} \\
\end{array} \right) \\
~~\hat{0} & ~~\hat{0} & ~~\hat{0} & 
~~\left( \begin{array}{cc}
X_+^{77} & X_+^{78} \\
X_+^{87} & X_+^{88} \\
\end{array} \right) \\
\end{array} \right) 
\end{equation}

Substituting these matrices into Eqs. (\ref{def:Xminus}) and 
(\ref{def:Xplus}), we can reduce the set to solving two quadratic 
matrix equations for the $2\times 2$ complex matrices 
$\hat{x}_{-}^{33}$ 
and $\hat{x}_{+}^{44}$. 
\begin{equation}
\hat{x}_{-}^{33}=
\left( \begin{array}{cc}
X_-^{55} & X_-^{56} \\
X_-^{65} & X_-^{66} \\
\end{array} \right) 
\nonumber
\end{equation}
and 
\begin{equation}
\hat{x}_{+}^{44}=
\left( \begin{array}{cc}
X_+^{77} & X_+^{78} \\
X_+^{87} & X_+^{88} \\
\end{array} \right). 
\end{equation}
This appears to be equivalent to solving a system of 8 simultaneous 
equations with real coefficients. Mathematically, one can 
never predict the number of solutions. We solve this system numerically, 
by applying Newton's method, which requires an initial guess of the solution. 
Since we know that the solution is close to the Andreev approximation, 
we give as an initial guess a diagonal matrix, namely the unitary matrix. 
This leads us to the physical solutions for $\hat{\mathbf X}_-$ and 
$\hat{\mathbf X}_+$ which, by using equation (\ref{def:Tjj}), allows the 
calculation of the $\hat{\mathbf T}_0$ matrix, which we need for the 
$\hat{\mathbf T}_{jj'}$ matrix (\ref{cs:Tml5}) and the Green's 
function (\ref{cs:Gmi3}). 
Finally, we are able to make use of the expressions (\ref{cs:dos}) 
and (\ref{cs:current}) and calculate the LDOS and the supercurrent of a 
periodic SN multilayer. 

In Andreev approximation some matrix elements are zero and 
further simplifications can be made. 
By neglecting 
the ordinary reflections of the quasiparticles at the SN interfaces, 
some of the $\hat{t}$-matrices are equal to zero. This results 
in a more simple 
form for the matrices $\hat{\mathbf A}$ and $\hat{\mathbf C}$ 
\begin{equation}
\hat{\mathbf A}=
\left( \begin{array}{cccc}
~~0 & ~~0 & ~~0 & ~~0 \\
~~0 & ~~0 & ~~0 & ~~\hat{a}_{24} \\
~~0 & ~~0 & ~~0 & ~~0 \\
~~0 & ~~0 & ~~0 & ~~0 \\
\end{array} \right), 
\end{equation}
where 
\begin{eqnarray}
\hat{a}_{24}&=&
\Biggl( \begin{array}{cc}
~~\hat{t}_{-+}^{+++-}d^+_{+j}e^{-ik_{+j}^+a_+} 
& ~~\hat{t}_{-+}^{+-+-}d^-_{+j}e^{ik^{-}_{+j}a_+} \\
~~\hat{t}_{-+}^{-++-}d_{+j}^+e^{-ik_{+j}^+a_+} 
& ~~\hat{t}_{-+}^{--+-}d_{+j}^-e^{ik_{+j}^-a_+} \\
\end{array} \Biggr)\nonumber \\
&=&\Biggl( \begin{array}{cc}
~~\hat{t}_{-+}^{+++-}d^+_{+j}e^{-ik_{+j}^+a_+} & ~~0 \\
~~0 & ~~\hat{t}_{-+}^{--+-}d_{+j}^-e^{ik_{+j}^-a_+} \\
\end{array} \Biggr). 
\end{eqnarray}

\begin{equation}
\hat{\mathbf C}=
\left( \begin{array}{cccc}
~~0 & ~~0 & ~~\hat{c}_{13} & ~~0 \\
~~0 & ~~0 & ~~0 & ~~0 \\
~~0 & ~~0 & ~~0 & ~~0 \\
~~0 & ~~0 & ~~0 & ~~0 \\
\end{array} \right), 
\end{equation}
where 
\begin{eqnarray}
\hat{c}_{13}&=&
\Biggl( \begin{array}{cccccccc}
~~\hat{t}_{+-}^{++-+}d_{-j}^+e^{-ik_{-j}^+a_-} 
& ~~\hat{t}_{+-}^{+--+}d_{-j}^-e^{ik_{-j}^-a_-} \\
~~\hat{t}_{+-}^{-+-+}d_{-j}^+e^{ik_{-j}^+a_-} 
& ~~\hat{t}_{+-}^{---+}d_{-j}^-e^{-ik_{-j}^-a_-} \\
\end{array} \Biggr)\nonumber \\
&=&\Biggl( \begin{array}{cccccccc}
~~\hat{t}_{+-}^{++-+}d_{-j}^+e^{-ik_{-j}^+a_-} & ~~0  \\
~~0 & ~~\hat{t}_{+-}^{---+}d_{-j}^-e^{-ik_{-j}^-a_-} \\
\end{array} \Biggr). 
\end{eqnarray}

By consequence, the solutions 
$\hat{x}_{-}^{33}$ and $\hat{x}_{-}^{44}$ get then a diagonal form. 
\begin{equation}
\hat{x}_{-}^{33}=
\left( \begin{array}{cccccccc}
X_-^{55} & 0 \\
0 & X_-^{66} \\
\end{array} \right) 
\nonumber
\end{equation}
and 
\begin{equation}
\hat{x}_{+}^{44}=
\left( \begin{array}{cc}
X_+^{77} & 0 \\
0 & X_+^{88} \\
\end{array} \right) 
\nonumber
\end{equation}

This allows us to decouple the equations for $X_-^{55}$ and $X_-^{66}$ into 
two quadratic equations which can be solved directly. The same holds for 
$\hat{x}^{44}_{+}$. Combining the two possible solutions 
for $X_-^{55}$ and 
$X_-^{66}$ with the two possible solutions for $X_+^{77}$ and $X_+^{88}$, 
one obtains four mathematical solutions for $\hat{\mathbf X}_{-}$ and 
$\hat{\mathbf X}_{+}$. Two of the solutions are complementary and lead to a 
zero value for the LDOS. Using the other two solutions, one gets either 
the positive physical value for the LDOS, or the same value with 
the opposite sign. We use this criterion to distinguish the physical 
solution from the four mathematically possible solutions. 

\end{appendix}

\newpage 

\begin{figure}[htbp]
\centerline{\epsfig{figure=DeltaTLt30.eps,angle=0,width=7.1cm}}
\caption[]
{Selfconsistent gap calculations for an homogeneous S bar of 
$L_t$=30 Bohr, using momenta cut-off. } 
\label{DeltaTLt30}
\end{figure}

\begin{figure}[htbp]
\centerline{\epsfig{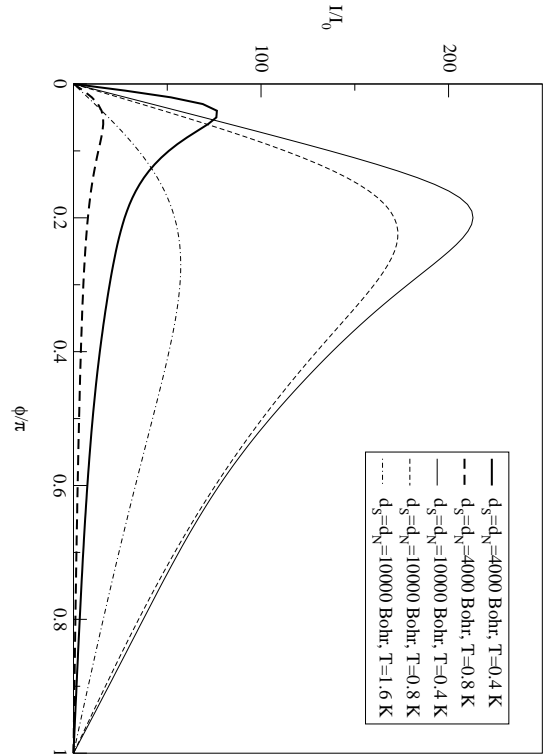}}
\caption[]
{Phase dependence of the supercurrent $I$, calculated with 
a selfconsistent gap function. $L_t=30$ and $I_0=e\Delta^s/\hbar$, with 
$\Delta^s=0.8\times10^{-4}$ the selfconsistent gap for the homogeneous 
bar at $T=0$. } 
\label{StroomNb}
\end{figure}

\begin{figure}[htbp]
\centerline{\epsfig{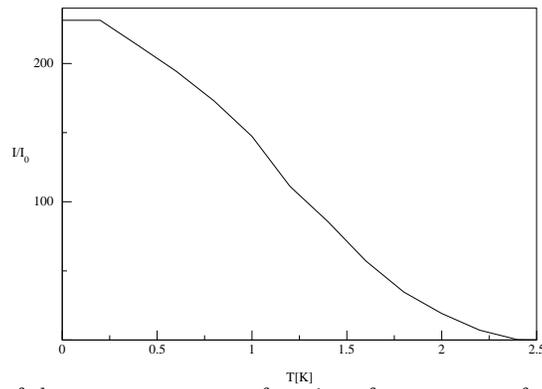}}
\caption[]
{Selfconsistent calculation of the supercurrent as a function of 
temperature for a SN multilayer with $d_S=d_N$=10000 and $L_t$=30. 
$I_0=e\Delta^s/\hbar$ and $\Delta^s=0.8\times10^{-4}$. } 
\label{it}
\end{figure}

\begin{figure}[htbp]
\centerline{\epsfig{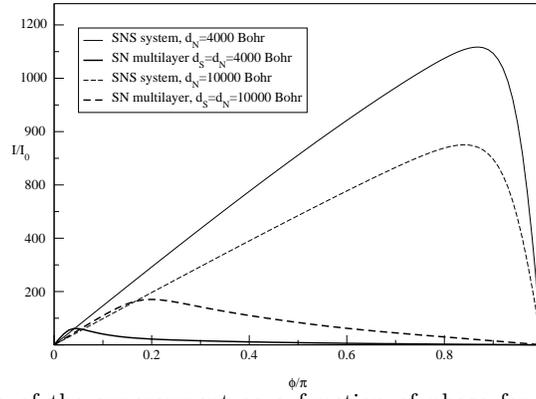}}
\caption[]
{Selfconsistent calculation of the supercurrent as a function of 
phase for SN multilayers with $d_S=d_N$=10000 and $d_S=d_N$=4000, 
and corresponding SNS systems with $d_N$=10000 and $d_N$=4000 respectively. 
$L_t$=30, $I_0=e\Delta^s/\hbar$, and $\Delta^s=0.8\times10^{-4}$. } 
\label{StroomSNS}
\end{figure}

\begin{figure}[htbp]
\centerline{\epsfig{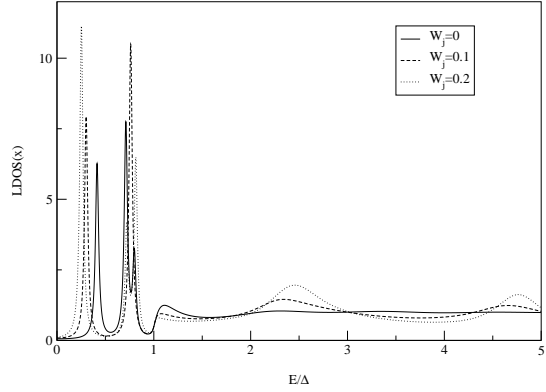}}
\caption[]
{LDOS for a SNS system at different barrier strengths $W$. 
$d_N=10000$, $L_t=13$. }
\label{potintSNS}
\end{figure}

\begin{figure}[htbp]
\centerline{\epsfig{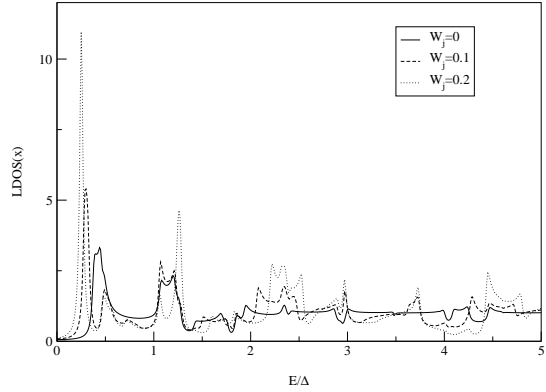}}
\caption[]
{LDOS for a SN multilayer at different barrier strengths $W$. 
$d_S=d_N=10000$, $L_t=13$. }
\label{potintLDOS}
\end{figure}

\begin{figure}[htbp]
\centerline{\epsfig{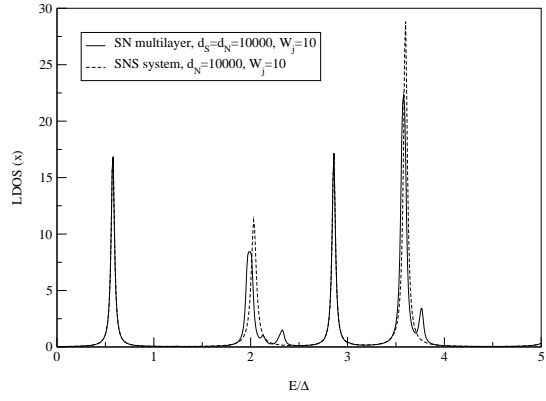}}
\caption[]
{LDOS in the N-layer of a SN multilayer and of a SNS system, 
at $W=10$ RydBohr. $d_S=d_N=10000$, $L_t=13$. }
\label{W10}
\end{figure}

\begin{figure}[htbp]
\centerline{\epsfig{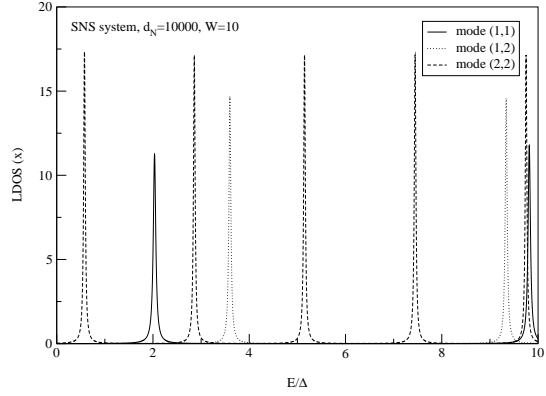}}
\caption[]
{The contributions of the transverse modes to the LDOS of a SNS system with 
$W=10$. $d_N=10000$, $L_t=13$. }
\label{W10modes}
\end{figure}

\begin{figure}[htbp]
\centerline{\epsfig{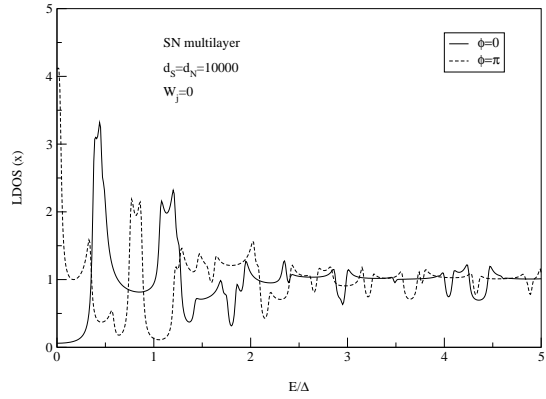}}
\caption[]
{LDOS in the N-layer of a SN multilayer at $\phi=0$ and $\pi$, and without 
interface barrier. $d_S=d_N=10000$, $L_t=13$. }
\label{W0ph}
\end{figure}

\begin{figure}[htbp]
\centerline{\epsfig{figure=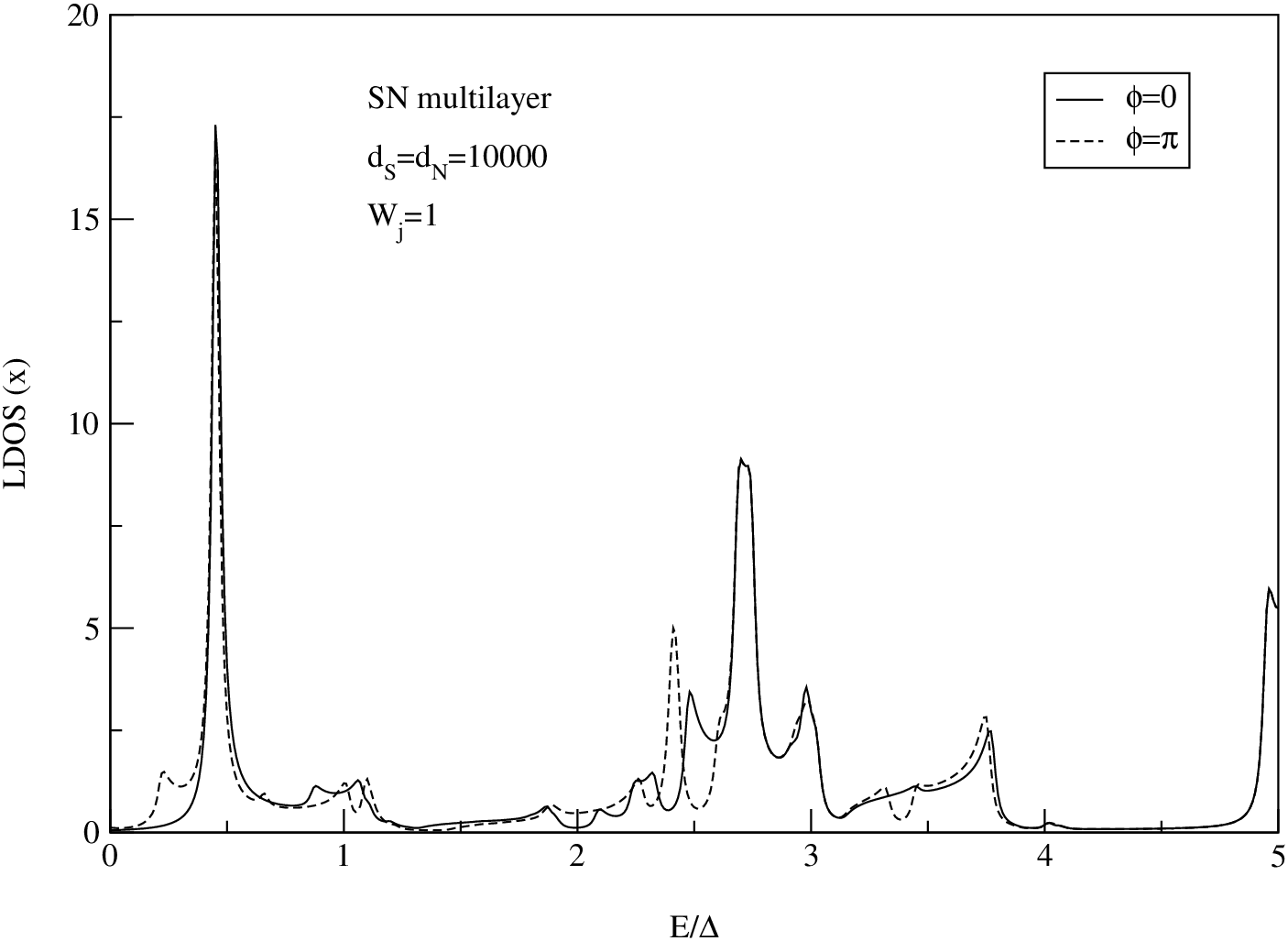,angle=0,width=7.1cm}}
\caption[]
{LDOS in the N-layer of a SN multilayer at $\phi=0$ and $\pi$, and with 
$W=1$. $d_S=d_N=10000$, $L_t=13$. }
\label{W1ph}
\end{figure}

\begin{figure}[htbp]
\centerline{\epsfig{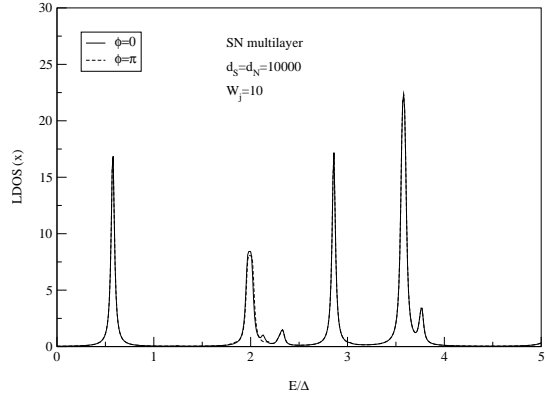}}
\caption[]
{LDOS in the N-layer of a SN multilayer at $\phi=0$ and $\pi$, and with 
$W=10$. $d_S=d_N=10000$, $L_t=13$. }
\label{W10ph}
\end{figure}

\begin{figure}[htbp]
\centerline{\epsfig{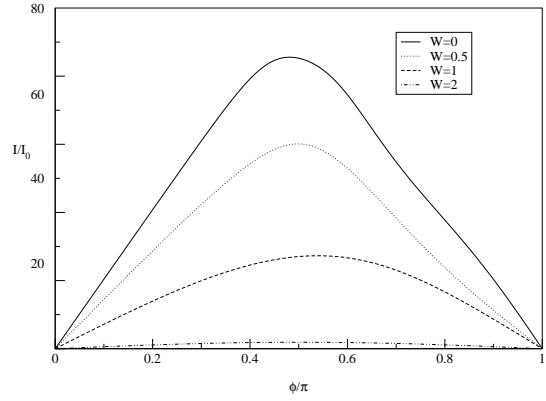}}
\caption[]
{Supercurrent for a SN multilayer at different potentials. 
$d_N=10000$, $L_t=13$, and $I_0=e\Delta/\hbar$. } 
\label{potintI}
\end{figure}


\newpage 

\begin{references}
\bibitem{mcmillan} W.L. McMillan and M. Rowell, in {\em Superconductivity}, 
        edited by R.D. Parks (Marcel Dekker, New York, 1969), p. 561; 
        W.L. McMillan, {\em Phys. Rev.} {\bf 175}, 537 (1968); 
        W.L. McMillan, {\em Phys. Rev.} {\bf 175}, 559 (1968). 
\bibitem{ishii} C. Ishii, {\em Prog. Theor. Phys} {\bf 47}, 1464 (1972),
        {\bf 44}, 1525 (1970). 
\bibitem{furusaki1} A. Furusaki and M. Tsukada, {\em Solid State Commun.} 
        {\bf 78}, 299 (1991); 
        A. Furusaki and M. Tsukada, {\em Physica B} {\bf 165 \& 166}, 
        967 (1990). 
\bibitem{furusaki} A. Furusaki, H. Takayanagi, and M. Tsukada, 
        {\em Phys. Rev. B} {\bf 45}, 10563 (1992). 
\bibitem{roland-th} R.T.W. Koperdraad, R.E.S. Otadoy, M. Blaauboer, and 
	A. Lodder, J. Phys.: Condensed Matter {\bf 13}, 8707 (2001). 
\bibitem{miriam}M. Blaauboer, R. T. W. Koperdraad, A. Lodder, 
	and D. Lenstra, Phys. Rev. B {\bf 54}, 4283 (1996).
\bibitem{roland2}R.E.S. Otadoy, and A. Lodder, Phys. Rev. B {\bf 65}, 
	024521 (2002). 
\bibitem{6:Tanaka91} Y. Tanaka and M. Tsukada,
	Phys.\ Rev.\ B {\bf 44}, 7578 (1991).
\bibitem{stojkovic}B. P. Stojkovic and O. T. Valls, Phys. Rev. B 
	{\bf 50}, 3374 (1994). 
\bibitem{miller}P. Miller and J. K. Freericks, J. Phys.: Condens. 
	Matter {\bf 13}, 3187 (2001). 
\bibitem{beenakker}C.W.J. Beenakker, Phys. Rev. B {\bf 67}, 3836 (1991). 
\bibitem{lesovik}N.M. Chtchelkatchev, G.B. Lesovik, and G. Blatter, 
	Phys. Rev. B {\bf 62}, 3559 (2000). 
\bibitem{blonder} G.E. Blonder, M. Tinkham, and T.M. Klapwijk, 
	Phys. Rev. B {\bf 25}, 4515 (1982). 
\bibitem{zaikin}A.V. Galaktionov and A.D. Zaikin, 
	Phys. Rev. B {\bf 65}, 184507 (2002). 
\bibitem{muller}R. Kleiner and P. M\"{u}ller, {\em Physica C} 
        {\bf 293}, 156 (1997). 
\end{references}
\end{document}